\documentclass[10pt, journal, compsoc]{IEEEtran}
\IEEEoverridecommandlockouts
\usepackage{cite}
\usepackage{amsmath,amssymb,amsfonts}
\usepackage{algorithm}
\usepackage[noend]{algpseudocode}
\usepackage{balance} 
\usepackage{fancyvrb}
\usepackage{graphicx}
\usepackage{subfig}[caption=false]
\usepackage{textcomp}
\usepackage{threeparttable}
\usepackage{verbatim}
\usepackage{xcolor}
\def\BibTeX{{\rm B\kern-.05em{\sc i\kern-.025em b}\kern-.08em
	T\kern-.1667em\lower.7ex\hbox{E}\kern-.125emX}}

\graphicspath{{figure/}}

\newcommand{\GW}{\textit{GW}}
\newcommand{\MB}{\textit{MB}}

\newcommand{\PRF}{\mathsf{PRF}}
\newcommand{\Sym}{\mathsf{Sym}}
\newcommand{\SHVE}{\mathsf{SHVE}}
\newcommand{\MSHVE}{\mathsf{SHVE}\text{+}}
\newcommand{\Adv}{\mathcal{A}}
\newcommand{\Sim}{\mathcal{S}}
\renewcommand{\P}{\mathcal{P}}
\newcommand{\Real}{\mathbf{Real}}
\newcommand{\Ideal}{\mathbf{Ideal}}

\newtheorem{theorem}{Theorem}

\begin{document}

\title{Practical Encrypted Network Traffic Pattern Matching for Secure Middleboxes}

\author{Shangqi Lai, Xingliang Yuan, Shi-Feng Sun, Joseph K. Liu, Ron Steinfeld, Amin Sakzad, Dongxi Liu
\IEEEcompsocitemizethanks{\IEEEcompsocthanksitem Shangqi Lai, Xingliang Yuan, Shi-Feng Sun, Joseph K. Liu, Ron Steinfeld and Amin Sakzad are with the Faculty of Information Technology, Monash University, Australia.\protect\\
Email: \{shangqi.lai,xingliang.yuan,shifeng.sun,joseph.liu,ron.steinfeld,\protect\\ amin.sakzad\}@monash.edu
\IEEEcompsocthanksitem Dongxi Liu is with Data 61, CSIRO, Australia.\protect\\
Email: dongxi.liu@data61.csiro.au
\IEEEcompsocthanksitem Copyright (c) 2021 IEEE. Personal use of this material is permitted. However, permission to use this material for any other purposes must be obtained from the IEEE by sending a request to pubs-permissions@ieee.org. Citation information: DOI 10.1109/TDSC.2021.3065652
}
}

\IEEEtitleabstractindextext{%
\begin{abstract}
	Network Function Virtualisation (NFV) advances the adoption of composable software middleboxes. Accordingly, cloud data centres become major NFV vendors for enterprise traffic processing. Due to the privacy concern of traffic redirection to the cloud, secure middlebox systems (e.g., BlindBox) draw much attention; they can process encrypted packets against encrypted rules directly. However, most of the existing systems supporting pattern matching based network functions require the enterprise gateway to tokenise packet payloads via sliding windows. Such tokenisation induces a considerable communication overhead, which can be over 100$\times$ to the packet size. 
	To overcome this bottleneck, in this paper, we propose the first bandwidth-efficient encrypted pattern matching protocol for secure middleboxes. We resort to a primitive called symmetric hidden vector encryption (SHVE), and propose a variant of it, aka SHVE+, to achieve constant and moderate communication cost. To speed up, we devise encrypted filters to reduce the number of accesses to SHVE+ during matching highly. We formalise the security of our proposed protocol and conduct comprehensive evaluations over real-world rulesets and traffic dumps. The results show that our design can inspect a packet over 20k rules within 100 $\mu$s. Compared to prior work, it brings a saving of $94\%$ in bandwidth consumption.
\end{abstract}
}

\maketitle

\section{Introduction}
Large-scale adoption of Network Function Virtualisation (NFV) facilitates easy realisation, deployment, and management of advanced network functions (aka middleboxes) for enterprises. Under this paradigm, cloud data centres become major NFV vendors~\cite{awsnfv17}. Traditionally dedicated and tightly coupled hardware/software is transformed into composable software middlebox modules, which can run on commodity cloud instances with unlimited scalability.
Such a technology shift also raises crucial privacy concerns, because the traffic of enterprises is re-directed and exposed to cloud data centres~\cite{wang2017toward,sultana2016light}. Although HTTPS is widely adopted, commercial middlebox services intercept and decrypt the traffic to retain advanced network functions like deep packet inspection (DPI)~\cite{SherryHS12}. 

To address this privacy concern, privacy-preserving middleboxes~\cite{sherry2015blindbox,yuan2016privacy,lan2016embark,fan2017spabox,ning2019privdpi,ning2020pine,poddar2018safebricks,duan2019lightbox,han2017sgx,trach2018shieldbox} have received much attention; these middleboxes are designed to process encrypted traffic against encrypted rules without decryption. 
%
As a result, sensitive traffic payloads and proprietary middlebox rules are protected without sacrificing the underlying operations of network functions, such as pattern matching, header inspection, and regular expression. 
Existing studies in this field can be classified into two categories, i.e., software-based solutions~\cite{sherry2015blindbox,yuan2016privacy,lan2016embark,fan2017spabox,ning2019privdpi,ning2020pine} and hardware-based solutions~\cite{poddar2018safebricks,duan2019lightbox,han2017sgx,trach2018shieldbox}. 
Unfortunately, neither of them are practically deployable due to efficiency and/or security issues. 


{\bf\noindent Limitations of prior work.} Mainstream software-based solutions~\cite{sherry2015blindbox,yuan2016privacy,lan2016embark,fan2017spabox, ning2019privdpi,ning2020pine} adapt a cryptographic technique named \emph{searchable encryption}~\cite{CurtmolaGKO11}, which allows middleboxes to match encrypted patterns extracted from rules against encrypted string streams (i.e., tokens) parsed from traffic payloads.
Those designs are communication inefficient, because traffic payloads need to be tokenised into string streams via sliding windows in varied sizes (i.e., enumerating the sizes of all specified patterns). 
As shown in prior work~\cite{sherry2015blindbox,yuan2016privacy,Desmoulins2018,ning2019privdpi,ning2020pine}, such cost can be tens of times to the original packet size. 
Consequently, long latency is inevitably introduced in token transmission, which is not acceptable in most networked applications. Besides, high I/O consumption between the enterprise and cloud greatly increases the capital cost of data transfer.  

Hardware-based solutions~\cite{poddar2018safebricks,duan2019lightbox,han2017sgx,trach2018shieldbox} rely on hardware enclave (i.e., Intel SGX) to execute middlebox functions in a trusted environment, in which traffic is fed into the enclave and processed within it. Using SGX brings benefits on efficiency and functionality for secure middleboxes, but side-channel attacks against SGX~\cite{BulckForeshadow18,van2019ridl,kocher2019spectre} make such adoption questionable. 

{\bf\noindent Contributions and technical overview.} To tackle the above limitations, in this paper, we aim to propose practical cryptographic protocols for a wide range of pattern matching-based secure middleboxes. Our design expects to offer convincing performance in both time and communication towards network environments, while ensuring cryptographic protection for rules and traffic payloads.

As mentioned, existing designs based on searchable encryption fall short of achieving bandwidth efficiency. 
To overcome this bottleneck, we observe that a cryptographic primitive named hidden vector encryption (HVE)~\cite{IovinoP08} can be a starting point of building a  bandwidth-efficient protocol. 
Specifically, HVE generates the key and ciphertext from two same-sized vectors, respectively, and HVE decryption can be performed only if the non-wildcard positions of the two vectors are the same.
In our context, the packet payload is encrypted from a vector of payload byte stream ($b_1, ..., b_n$), while the rule pattern is encrypted from a $n$-byte predicate vector ($*, \ldots, *$, $p_1, ..., p_m$, $*, \ldots, *$) with wildcard ($*$) positions. 
Later, the middlebox performs HVE decryption on those two vectors to check if there is a match. 
With HVE, the communication overhead becomes constant, as traffic is encrypted in byte-wise.
%
For efficiency, we exclude public-key HVE schemes and resort to the symmetric-key HVE scheme, aka SHVE~\cite{lai2018result}. 
%
%
%

%
%
%

The original SHVE scheme~\cite{lai2018result} is designed for encrypted membership testing only, where the message is not embedded in the SHVE ciphertext. Thus, it cannot be directly applied to pattern matching-based middlebox functions like DPI, because a DPI rule contains patterns and the corresponding action  (e.g., alert, drop), and the entire rule should fully be protected without matching~\cite{asghar2016splitbox,yuan2016privacy}. 
To solve this problem, we slightly modify the construction of SHVE and propose a variant of it called SHVE+ which supports both message encryption and byte-wise encrypted matching. 
Our primitive provides the same security guarantees of the existing secure middlebox systems~\cite{sherry2015blindbox,yuan2016privacy,ning2019privdpi,ning2020pine}. The equality of byte strings in a packet payload is fully hidden, and the action is triggered only if a match is found in the encrypted payload at the positions specified in DPI rules.  


To improve efficiency, we design a fine-grained progress filtering protocol to reduce the number of accesses on SHVE+ during the matching process. Our key insight is that most of the packets are commonly legitimate~\cite{choi2016dfc, yuan2016privacy, stylianopoulos2017multiple}, and if a packet is filtered out, i.e., being identified as a mismatch, the middlebox will stop processing it to save processing time. 
%
%
%
%
To apply filtering to encrypted traffic, we propose an encrypted filter structure via SHVE. 
%
%
It is carefully designed in a way that the encrypted packet payload can be used for both filtering and pattern matching. Namely, introducing our filters does not incur extra bandwidth cost. 

For completeness, we formalise the security of our proposed protocols. First, we formally capture the capabilities of adversaries considered in the targeted middlebox system. 
%
One adversarial model aims to infer sensitive information from encrypted packets, while the other aims to deduce information from encrypted rulesets.
%
%
%
We note that security analysis of existing designs~\cite{sherry2015blindbox, fan2017spabox, yuan2016privacy, guo2018enabling,ning2019privdpi} does not appear to capture both adversaries at the same time.  
%
To bridge the gap, we adapt the real/ideal paradigm to define two groups of games under the above two adversarial models.
We prove that even if an adversary is capable of selecting the packet or ruleset to be challenged in advance, she only learns a controlled leakage profile regarding the packet and ruleset.

We implement a prototype and deploy it on a commodity machine. We use real-world patterns (Snort and ETOpen) and network traffic (iCTF08) to evaluate its performance.
Regarding latency, our middlebox can inspect a packet for ETOpen ruleset (20k+ rules) within $100~\mu$s and Snort ruleset (1.5k rules) within $60~\mu$s. 
In a multi-session scenario ($100$ concurrent connections), the throughput per connection reaches $5000$ packets per second for the Snort ruleset and $3000$ packets per second for the ETOpen ruleset.
The overall throughput is over $1$ GBps and $500$ MBps, respectively.
Regarding bandwidth consumption, our design consumes the least bandwidth among all prior arts (including the one in~\cite{Desmoulins2018} also with constant complexity): it only costs 5 times more bandwidth in terms of the original packet size, which saves more than $94\%$ comparing the designs~\cite{sherry2015blindbox,lan2016embark,yuan2016privacy,ning2019privdpi,ning2020pine} using tokenisation.
%
The cost estimation based on AWS pricing information demonstrates that the monthly maintenance cost of our middlebox is $\$460.8$, which is only one-fourth of the tokenisation-based designs.

Our contributions can be summarised as follows:
\begin{itemize}
	\item We design a variant of the SHVE scheme called SHVE+, which preserves the functionality, efficiency, and security properties of SHVE while additionally supporting message encryption.
	\item We propose the first bandwidth-efficient encrypted pattern matching protocol built from SHVE+, which enables middleboxes to perform pattern matching over encrypted traffic with constant and moderate bandwidth overhead. 
	\item We propose a secure filter to filter out legitimate packets, which further improves the efficiency of our design by $5\times$ to $8\times$.
	Meanwhile, this optimisation does not incur extra bandwidth cost as it reuses the encrypted traffic for pattern matching.
	\item We are the first to comprehensively formalise the security of encrypted pattern matching protocols for secure middleboxes. 
	We formally prove that our protocol protects against the adversary who wants to compromise traffic and rules throughout pattern matching.
	%
	\item We implement a system prototype and evaluate it with real-world rulesets and a traffic dump. 
	We evaluate the setup time, memory cost, inspection delay, bandwidth overhead, throughput, and deployment cost of our system, and compare them with two prior encrypted pattern matching protocols (i.e., BlindBox~\cite{sherry2015blindbox} and SEST~\cite{Desmoulins2018}).
\end{itemize}

\section{Related Work}
{\bf\noindent Software-based secure middleboxes.} Our work is related to software-based (aka cryptographic) solutions for secure middleboxes. Blindbox~\cite{sherry2015blindbox} is the first system that supports pattern matching based network functions over the encrypted traffic. It is also the first to use searchable encryption for encrypted pattern matching. Later, a line of work is proposed to improve the design of Blindbox, including the realisation of rule re-using~\cite{ning2019privdpi, ning2020pine}, header matching~\cite{lan2016embark,guo2018enabling,guo2020privacy}, dedicated inspection rule~\cite{yuan2016privacy} and regular expression~\cite{fan2017spabox}. As mentioned, these designs built from searchable encryption require tokenisation of the payloads, which is a critical performance bottleneck of the system (can lead $77$ -- $120\times$ bandwidth overhead in terms of the original traffic size). 
Some other studies built secure middleboxes from advanced cryptographic tools such as secure multi-party computation~\cite{asghar2016splitbox,lai2019enabling} and public-key searchable encryption~\cite{canard2017blindids}.
Although they support secure anomaly matching/detection, they are not communication efficient either.

{\bf\noindent Hardware-based secure middleboxes.} There are also solutions based on trusted hardware, i.e., Intel SGX. These designs~\cite{poddar2018safebricks,duan2019lightbox,trach2018shieldbox,han2017sgx} aim to achieve the same goal of processing encrypted traffic, yet using trusted hardware enclave. As mentioned, Intel SGX is currently vulnerable to side-channel attacks~\cite{BulckForeshadow18,van2017telling,kocher2019spectre}, which can break the security guarantee of the trusted enclave. 
Although oblivious primitives~\cite{lai2021oblivsketch} can help mitigate some of those attacks (e.g.,\cite{van2017telling}), it is not a panacea for all side-channels.
Moreover, deploying those systems requires the cloud servers to be equipped with SGX and enforces the enterprises to trust the hardware vendor. 
The above constraints would limit the adoption of these SGX-based systems.

{\bf\noindent Pattern matching on encrypted data.} In the literature, some theoretical work also investigates pattern matching on encrypted data. 
A recent scheme~\cite{Desmoulins2018} based on cryptographic pairing achieves a constant communication overhead to the packet size. 
Also, it supports secure pattern matching in a multi-session environment since it is based on the public-key encryption scheme.
Unfortunately, such a theoretical design still introduces unaffordable bandwidth overhead in practice, i.e., $64\times$ larger than the original traffic. Besides, the pairing based matching operation is too slow to be deployed in traffic processing. 
A detailed comparison can be found in Section~\ref{sec:eva}.

Some early studies are working on substring matching~\cite{chase2015substring,hahn2018practical}. Those studies focus on different application scenarios, where a long string is encrypted and stored at the server, and later a substring (pattern) query will be issued to be processed against the long string for matching. 

\begin{table}
	\centering
	\caption{Summary of the performance of representative software-based secure pattern matching middleboxes.}
	\label{tbl:comparison}
	\tabcolsep 0.02in
	\begin{tabular}{cccc}
		\hline
		Scheme & Communication & Memory cost & Inspection delay \\
		\hline 
		SEST~\cite{Desmoulins2018} & High & Medium & ms level\\
		\hline
		Splitbox~\cite{asghar2016splitbox} & High & Medium & ms level\\
		\hline
		Tokenisation~\cite{sherry2015blindbox, yuan2016privacy, lan2016embark, fan2017spabox,ning2019privdpi,ning2020pine} & High & Low & $\mu$s level\\
		\hline
		Our middlebox & Low & High & $\mu$s level\\
		\hline
	\end{tabular}
\end{table}

To summarise, we present a comparison table (Table~\ref{tbl:comparison}). 
It shows that our proposed design outperforms the existing cryptographic works~\cite{asghar2016splitbox,Desmoulins2018} in terms of the communication cost and inspection delay.
It highly reduces the communication cost in tokenisation-based approaches~\cite{sherry2015blindbox, yuan2016privacy, lan2016embark, fan2017spabox, ning2019privdpi,ning2020pine} while preserving a microsecond-level inspection delay.
Although its memory cost is higher than the other solutions, it is not an issue for in-cloud middleboxes. 
%
%
In networked applications, latency is crucial to user experience and quality of service~\cite{SherryHS12,todd2009latency}. The latency of traffic processing is more sensitive to bandwidth while rule encryption and upload is one-time setup cost. 
%
%
Besides, bandwidth is much more expensive than memory in the modern cloud (see Section~\ref{sec:eva}). Our solution offers a significant maintenance cost saving in the real-world deployment.

\section{Overview}
\subsection{System Architecture}
\label{sec:arch}
Our proposed design employs the same architecture as existing secure middleboxes~\cite{yuan2016bringing,yuan2016privacy,lan2016embark,wang2017toward} (just to list a few); it redirects an enterprise's traffic from the enterprise gateway to a third-party middlebox service for pattern matching-based packet processing. 
During this process, the enterprise leverages the middlebox to thoroughly inspect all traffic and enforce its security rules to defend against malicious activities.
In addition, the enterprise aims to protect the ruleset in the outsourced environment, because this can either be proprietary ruleset subscribed from professional vendors~\cite{sherry2015blindbox} or customised open-sourced ruleset with private information~\cite{yuan2016privacy}, e.g., enterprise's trade secrets, or intellectual property.

\begin{figure}[!t]
	\centering
	\includegraphics[width=\linewidth]{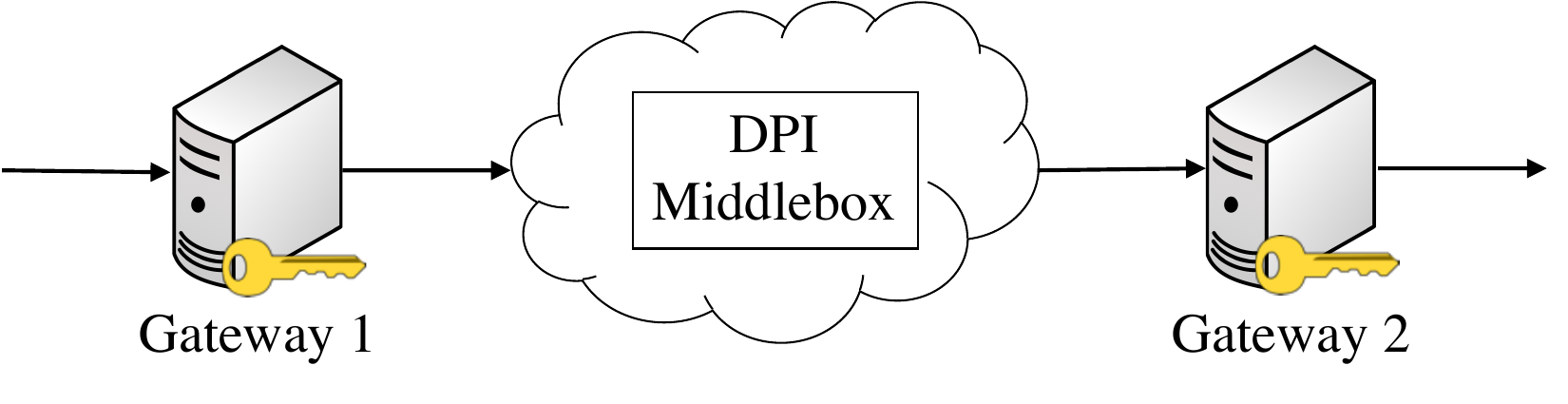}
	\caption{System Architecture. The arrows indicate traffic from the sender network to the receiver network; the response traffic follows the reverse direction.}
	\label{fig:architecture}
\end{figure}

Fig.~\ref{fig:architecture} presents the system architecture\footnote{If an enterprise endpoint connects to an external network, the processed traffic from the middlebox is sent back to the gateway, then sent out~\cite{yuan2016bringing,lan2016embark}.}.
It has two parties: the gateway (\GW) maintained by the enterprise and the middlebox (\MB) deployed in the  service provider, like public clouds. 
We also use the term ``endpoint'' to denote the server within the enterprise.
The system flow involves three phases:

{\bf\noindent Initialisation.} Before initiating any connection, \GW1~randomly chooses a key $msk$ and uses it to generate encrypted rules to be used by \MB~for detecting malicious packet payloads.
In practice, each rule describes an attack via its representative patterns, which may include suspicious strings in the payload and the offset information for the string~\cite{snort2019ruleset}. Each rule also indicates the corresponding actions (e.g., alert, drop) once a match is found. 
Thus, \GW1~creates an encrypted list for the pattern-action list extracted from the ruleset, which is later used for \MB~to match those strings in the encrypted payload and perform the associated action.
Meanwhile, \GW1~builds an encrypted filter, which can quickly process the mismatches in traffic, and it accelerates the pattern matching process.
The generated encrypted filter and pattern list are uploaded to \MB.
Later, \MB~can perform packet inspection for all incoming traffic through the encrypted filter and pattern list.

{\bf\noindent Preprocessing.} \GW1~should preprocess the packet payload before sending it for inspection.
Specifically, \GW1~scans the packet payload in byte-wise and uses $msk$ to generate encrypted traffic dedicated to the pattern matching service like DPI.
Then, \GW1~will send the encrypted traffic from the enterprise network to \MB.

{\bf\noindent Inspection.} Upon receiving the encrypted traffic, \MB~withholds the incoming traffic and executes the proposed encrypted pattern matching protocol to inspect the traffic with the pre-computed encrypted pattern list.
If an action can be recovered after checking the encrypted patterns, \MB~will apply the action to the packet; otherwise, the packet is considered as legitimate, and \MB~then sends it out to \GW2.
To improve the efficiency of the above process, \MB~exploits a secure filtering protocol.
In specific, for each packet, \MB~utilises the pre-built encrypted filter to quickly evaluate whether the current position in the encrypted traffic is a possible matching position.
As a result, \MB~separates the innocuous input from the possible malicious traffic, and it only runs the pattern matching protocol for those possible matching positions instead of checking the whole traffic with the encrypted patterns.

{\noindent \textbf{\textit{Remark}}.} Following prior studies~\cite{sherry2015blindbox, yuan2016privacy} that support pattern matching over the encrypted traffic, the real network traffic is protected by SSL (denoted as SSL traffic).
That is, \GW1 initialises a normal SSL connection with \GW2 and sends  SSL traffic with encrypted traffic to \MB~for inspection.
\GW2 can use its SSL session key to recover the real network traffic. 

\subsection{Threat Assumption}
\label{sec:threat}
We assume that \GW~in the enterprise network is a trustworthy party. It follows the proposed protocol and does not disclose the ruleset to other parties.
On the other hand, the \MB~service provider is assumed to be  semi-honest. 
It also follows the protocol to offer pattern matching service but attempts to extract sensitive data from the encrypted traffic passing through the middlebox and infer the private ruleset owned by the enterprise.
Also, the middlebox can be compromised or eavesdropped as it is deployed in an untrusted environment~\cite{sherry2015blindbox}.
Therefore, the main goal of the proposed system is to hide both the content of traffic and the ruleset from \MB~while allowing \MB~to perform pattern matching over the encrypted traffic.

We also assume that at least one endpoint in the communication is honest.
This is consistent with the threat model in existing privacy-preserving pattern matching middleboxes~\cite{sherry2015blindbox, yuan2016privacy,fan2017spabox, ning2019privdpi,ning2020pine}.
Note that detecting two malicious endpoints is an orthogonal work, and we do not consider this case in our paper.

\subsection{Building Blocks}
\label{sec:primitives}

{\bf\noindent Basic cryptographic tools.} We leverage pseudo-random function $\PRF$, which is a polynomial-time computable function family that is computationally indistinguishable from random functions to any probabilistic polynomial-time adversary.
Besides, we make use of symmetric key encryption scheme $\Sym$, which consists of three probabilistic polynomial-time algorithms $(\mathsf{KeyGen}, \mathsf{Enc}, \mathsf{Dec})$. $\mathsf{KeyGen}(\cdot)$ generates the secret key $k$. A message $m$ can be encrypted as a ciphertext $c\leftarrow \mathsf{Enc}(k, m)$ and decrypted by $m\leftarrow\mathsf{Dec}(k, c)$.
The formal definitions of the $\PRF$ and $\Sym$ can be found in~\cite{lai2018result}.

{\bf\noindent SHVE.} SHVE~\cite{lai2018result} is a predicate encryption scheme that supports conjunctive, equality, comparison and subset membership queries over the encrypted data.
Compared to the public-key HVE schemes~\cite{IovinoP08}, SHVE is much faster~\cite{lai2018result} as it only relies on the $\PRF$ and symmetric key encryption.
We present a brief definition of SHVE on below.

Let $\Sigma$ be an attribute set and $\ast$ be a wildcard symbol (``don't care'' value). We define $\Sigma_\ast = \Sigma \cup\{\ast\}$.
Let $\mathbf{x}=(x_1, ..., x_n)$ with $x_i\in\Sigma$ be an attribute vector, and $\mathbf{v}=(v_1, ..., v_n)$ with $v_i\in\Sigma_\ast$ be a predicate vector. The predicate function $P_{\bf v}({\bf x})=1$ if and only if for each $i\in[1, n]$, we have $x_i=v_i$ or $v_i=\ast$.
In other words, the predicate function returns ``$1$'' only when the vector $\mathbf{x}$ matches $\mathbf{v}$ in all non-wildcard positions.
The SHVE scheme uses a PRF $F_0$ and the symmetric key encryption $\Sym$ as described above. It comprises four probabilistic polynomial-time algorithms:
\begin{itemize}
	\item $\SHVE.\mathsf{Setup}(\lambda)$: On input the security parameter $\lambda$, the algorithm outputs the master secret key $msk\xleftarrow{\$}\{0, 1\}^\lambda$.
	\item $\SHVE.\mathsf{KeyGen}(msk, \mathbf{v})$: On input the master secret key $msk$ and a predicate vector $\mathbf{v}=(v_1, ..., v_n)$, the algorithm outputs a query trapdoor $\mathbf{s}= \left(d_0,d_1, S\right)$, where $d_0$ is a masked key, $d_1$ is a symmetric ciphertext and $S$ keeps all non-wildcard positions in $\mathbf{v}$.
	%
	\item $\SHVE.\mathsf{Enc}(msk, \mathbf{x})$: On input the master secret key $msk$ and an attribute vector $\mathbf{x}=(x_1, ..., x_n)$, this algorithm sets $c_l=F_0(msk, x_l||l)$ for each $l\in[n]$, and outputs the ciphertext $\mathbf{c}=(\{c_l\}_{l\in[n]})$.
	\item $\SHVE.\mathsf{Query}(\mathbf{s},\mathbf{c})$: The query algorithm takes as input a trapdoor $\mathbf{s}$ and a ciphertext $\mathbf{c}$. 
	If the algorithm recovers $0$ from $\mathbf{s}$ and $\mathbf{c}$, the query algorithm outputs ``True'' (indicating $P_{\bf v}({\bf x})=1$) else it outputs $\bot$.
\end{itemize}

\subsection{Technical Overview}
To enable secure and efficient pattern matching over the encrypted traffic, we design a customised SHVE scheme (SHVE+), which can embed an encrypted message into the original SHVE query trapdoor.
We then devise a cryptographic protocol based on SHVE+.
In the proposed protocol, \GW~treats the pattern of each rule as a byte array and uses it as the predicate vector to generate an SHVE query trapdoor.
The corresponding action is encrypted and embedded into the trapdoor.
\GW~leverages SHVE+ to generate the query trapdoors of all rules in a ruleset and gives them to \MB.
Then, \GW~scans the packet in byte-wise and takes it as the attribute vector to generate an SHVE ciphertext.
During the inspection phase, \GW~sends the ciphertext to \MB, and \MB~leverages each query trapdoor to query the ciphertext. 
\MB~can recover an action from the trapdoor and perform it only when the pattern exists in the ciphertext.
The proposed protocol only relies on symmetric primitives and enables byte-wise matching, which makes it more computation/communication efficient than prior works based on tokenisation~\cite{sherry2015blindbox, yuan2016privacy, lan2016embark, fan2017spabox, ning2019privdpi} or public-key encryption~\cite{Desmoulins2018}.
Moreover, it can protect both the ruleset and network traffic against adversaries who compromise \MB.

To facilitate the matching process further, we propose a secure filtering protocol. 
In this protocol, \GW~utilises SHVE to generate the first 2 to 4 bytes of each pattern as query trapdoors.
Before examining the encrypted traffic with the pattern matching protocol, \MB~leverages the above SHVE trapdoors to query the ciphertext.
Since the trapdoor contains the position information, the filtering protocol can determine whether the pattern exists in the traffic or not as well as the position of all possible matches.
Hence, the pattern matching protocol's workload is considerably reduced because most of the packets are innocuous~\cite{choi2016dfc}, and those packets will be filtered out by the secure filtering protocol (only 1/9 -- 1/6 packets remain for pattern matching as in Section~\ref{sec:eva}). 
Note that the ciphertext structure of SHVE is identical to the one for SHVE+.
As a result, the secure filtering protocol can effectively filter out the innocuous traffic while introducing no extra bandwidth cost.

\section{The Proposed System}
\subsection{Construction of SHVE+}
\label{sec:MSHVE}
SHVE~\cite{lai2018result} can be adapted to achieve efficient encrypted pattern matching for network traffic.
However, it cannot be directly used for middlebox functions like DPI. As mentioned in Section~\ref{sec:arch}, the inspection rule consists of the inspection patterns and corresponding action.
To fully protect the rules during the matching process, both of them should be encrypted.
Also, to preserve the functionality, the action needs to be recovered for \MB~further processing when the pattern is matched.
To this end, the action should be considered as a message encrypted with the pattern in SHVE.
Similar to the design in prior work~\cite{yuan2016privacy, asghar2016splitbox}, the above design encrypts both the rule and action to minimise the leakage in the ou tsourced middlebox.
Nonetheless, we also take the performance into consideration and choose to reveal the action for those matched patterns.
This trade-off enables our middlebox to efficiently and securely handle a large volume of packets at a moderate cost and well-defined leakage.
We note that the original SHVE construction~\cite{lai2018result} can only be used for membership testing, whereas the message encryption is yet to be supported.
To address this issue, this section presents a new SHVE scheme, dubbed SHVE+, which enables message encryption on SHVE.

{\bf\noindent Construction.} The original SHVE (see Section~\ref{sec:primitives}) leverages a random key $K$ to encrypt ``$0$'' in the SHVE trapdoor ($d_{1}$ in the trapdoor), and it refers to the predicate vector $\mathbf{v}$ to masks the random key and keeps the masked key in $d_0$. 
If $\mathbf{v}$ matches the attribute vector encrypted in the ciphertext $\mathbf{c}$ at all non-wildcard positions, the encrypted ``$0$'' can be recovered from the trapdoor after $\SHVE.\mathsf{Query}$, and SHVE outputs ``True''.
Intuitively, we can exploit the $d_{1}$ term in the trapdoor to store the other encrypted message. Then, $\MSHVE.\mathsf{Query}$ is changed to return the decryption of $d_{1}$ after decrypting $d_{1}$ successfully.

We now present the details of our SHVE+ construction.
Note that only the modified algorithms are given here, the other algorithms remain the same as in Section~\ref{sec:primitives}.
\begin{itemize}
	\item $\MSHVE.\mathsf{KeyGen}(msk, \mathbf{v}, m)$: On input the master secret key $msk$, a predicate vector $\mathbf{v}=(v_1, ..., v_n)$ and a message $m$, the algorithm extracts all non-wildcard positions $S=\{l\in[n]|v_l\neq\ast\}$ from $\mathbf{v}$. 
	Let these positions be $l_1<\ldots<l_{|S|}$, the algorithm samples $K\xleftarrow{\$}\{0, 1\}^\lambda$ and computes:
	$d_{0} = \oplus_{j\in[|S|]} \left(F_0(msk, v_{l_j}||{l_j})\right) \oplus K$, $d_{1} = \Sym.\mathsf{Enc}\left(K,m\right)$
	Finally, it outputs the trapdoor $\mathbf{s}= \left(d_0,d_1, S\right)$ corresponding to the predicate vector $\mathbf{v}$.
	\item $\MSHVE.\mathsf{Query}(\mathbf{s},\mathbf{c})$: The query algorithm takes as input a trapdoor $\mathbf{s}$ and a ciphertext $\mathbf{c}$. Then, it computes $K' = \left(\oplus_{j\in [|S|]}c_{l_j}\right) \oplus d_0$ and returns $ \mu=\Sym.\mathsf{Dec}\left(K',d_{1}\right)$.
\end{itemize}

Under the SHVE+ scheme, the proposed system can encrypt the pattern as a SHVE+ trapdoor and then encrypt traffic as the SHVE+ ciphertext to make the inspection.
In specific, \GW~in the proposed system generates a pattern array initialised with wildcard character `$*$' in all positions. Then it inserts each byte of the pattern string into the pattern array according to the rule (string content, start/end position), and uses the array as the predicate vector and the action as the message to compute the encrypted pattern via $\MSHVE.\mathsf{KeyGen}$.
Later, \GW~parses the traffic into a byte array and uses it as the attribute vector to get the encrypted traffic by $\MSHVE.\mathsf{Enc}$.
Finally, on \MB, the encrypted pattern can examine traffic in the form of SHVE+ ciphertext and properly recovers the action if a match is found according to the definition of SHVE~\cite{lai2018result}.

{\bf\noindent Security.} SHVE+ retains SHVE's security properties for membership testing, which guarantees that the pattern matching process only reveals whether the encrypted traffic includes the pattern in the position specified by rules but nothing more.
Moreover, it ensures that the message can only be recovered when the traffic matches the pattern, which is consistent with the security requirement of the proposed middlebox service.
A detailed analysis is given in Section~\ref{sec:security}.

\begin{figure}[!t]
	\centering
	\includegraphics[width=\linewidth]{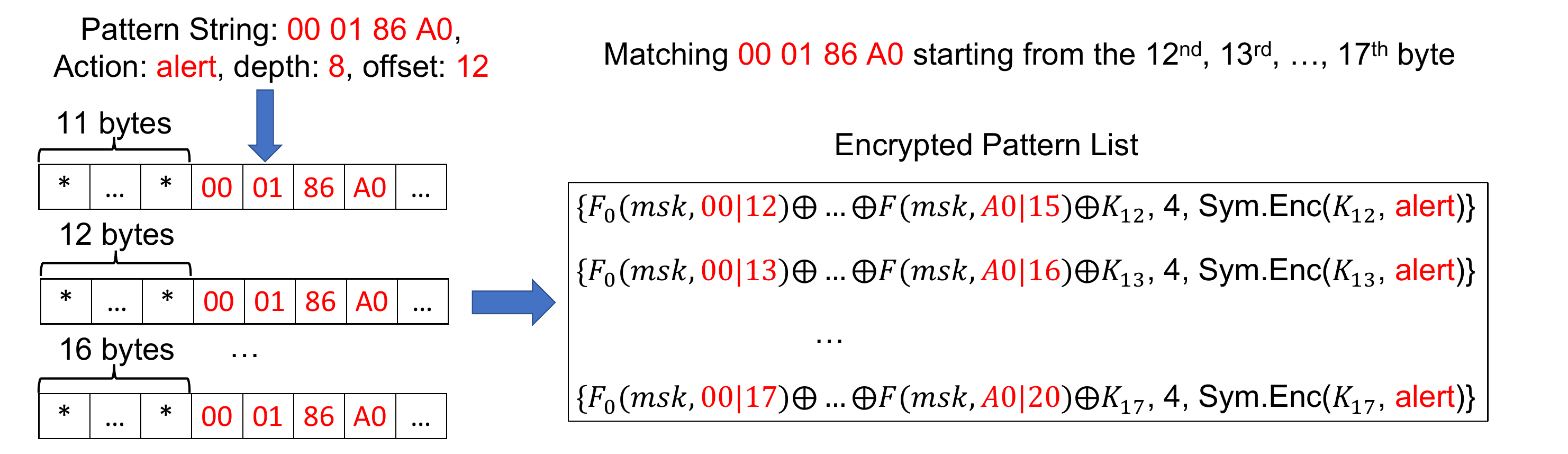}
	\caption{An example of the encrypted pattern generation: each pattern string is inserted into multiple matching arrays to match the pattern in every possible position in the payload.}
	\label{fig:predicatevector}
\end{figure}

\begin{algorithm}[!t]
	\caption{Encrypted Rule Generation}
	\label{Alg1}
	\small
	\begin{algorithmic}[1]
		\Require The master secret key $msk$ of SHVE+; the ruleset $\mathit{R}$
		\Ensure  The encrypted pattern list $\mathcal{E}$
		\Function {Generate}{$msk$, $\mathit{R}$}
		\State Parse $\mathit{R}$ as a pattern-action list $T=\{(pat, act)\}$
		\For{each $(pat, act)$ in $T$}
		\State $start\leftarrow pat.\textrm{offset}>0?pat.\textrm{offset}:1$
		\State $end\leftarrow pat.\textrm{depth}>0?start + pat.\textrm{depth}:1500$
		\For{$i=start : end - pat.\textrm{string}.len+1$}
		\State $\mathbf{v}_i\leftarrow *^{1500}$
		\State Insert $pat.\textrm{string}$ at $\mathbf{v}_i[i]$
		\State $\mathbf{t}_i\leftarrow\MSHVE.\mathsf{KeyGen}(msk, \mathbf{v}_i, act)$
		\State Store $\mathbf{t}_i$ in $\mathcal{E}$
		\EndFor
		\EndFor
		\State\Return $\mathcal{E}$
		\EndFunction
	\end{algorithmic}
\end{algorithm}
\begin{algorithm}[!t]
	\caption{Rule Matching}
	\label{Alg2}
	\small
	\begin{algorithmic}[1]
		\Require \GW~inputs the master secret key $msk$, the payload $P$; \MB~inputs the encrypted pattern list $\mathcal{E}$; 
		\Function {Match}{$msk, \mathcal{E}, P$}\\
		On \GW:
		\State Parse $P$ as a byte array and compute the encrypted traffic $\mathbf{c}\leftarrow\MSHVE.\mathsf{Enc}(msk, P)$
		\State Send $\mathbf{c}$ to \textit{MB}\\
		On \MB:
		\For{each encrypted pattern $\mathbf{t}$ in $\mathcal{E}$}
		\State $act'\leftarrow\MSHVE.\mathsf{Query}(\mathbf{t},\mathbf{c})$
		\State Execute $act'$ if it is valid
		\EndFor
		\EndFunction
	\end{algorithmic}
\end{algorithm}
\subsection{The Proposed Encrypted Pattern Matching Protocol}\label{sec:basic}
In order to support secure pattern matching over the encrypted traffic, the existing work~\cite{sherry2015blindbox, yuan2016privacy} leverages an encrypted index built from the pattern-action list. 
More specifically, the encrypted index is indexed by the encryption of each pattern string.
When a given inspection token matches the encrypted indexing term, \MB~can recover the action from the index and execute it.
However, due to the complexity of matching patterns (various size, matching position, etc.), this approach has to tokenise the original packet payload into a large number of tokens, and it can blow up the bandwidth consumption ($24\times$ as reported in~\cite{sherry2015blindbox}).
To enable pattern matching in a bandwidth-saving manner, our system is built from SHVE+ because it does not rely on any tokenise algorithm.
Instead, it encrypts the payload and queries the pattern in byte-wise.
Consequently, its bandwidth consumption is a constant no matter how long the pattern is (see Section~\ref{sec:MSHVE}).

{\bf\noindent Pattern matching for arbitrary pattern strings.} Algorithm~\ref{Alg1} summarises the detailed encrypted rule generation procedure run by \GW.
As mentioned, each inspection rule is parsed as a pattern-action tuple, and our protocol generates the encrypted pattern list from it.
In practice, the inspection rules often involve qualifiers that specific a range of positions to be checked in the packet payload. 
For example, the following Snort rule~\cite{snort2019ruleset} specifies the ``depth'' (only search 8 bytes instead of 1500 bytes for the pattern) and ``offset'' (start to search the pattern from the $12${th} byte of the payload).
\begin{Verbatim}[fontsize=\small,frame=single]
alert udp $EXTERNAL_NET any -> $HOME_NET 111 
(flow:to_server; content:"|00 01 86 A0|",
depth 8,offset 12)
\end{Verbatim}
\vspace{-5pt}
Thus, our protocol takes the above two qualifiers into consideration when generating pattern arrays and encrypted patterns.
%
As shown in Fig.~\ref{fig:predicatevector}, our protocol first generates $\text{offset}-\text{string}.len+1$ pattern arrays with wildcard character `$*$'. 
Then, it inserts the pattern string into the pattern arrays at each possible starting positions, which is 12 to 17 in our example.
For each pattern array, our protocol inputs the action and runs $\MSHVE.\mathsf{KeyGen}$ to encrypt the pattern string and position after concatenating them together (see Section~\ref{sec:MSHVE}).
This ensures that matches only happen on the positions specified by the rule.
The result encrypted pattern list is able to match the pattern in the specific positions over any incoming traffic from $\GW$.

The above protocol supports encrypted pattern matching in wildcard positions, i.e., the pattern can be found in all positions in a packet.
For this case, our protocol generates encrypted patterns for all positions to find matches in traffic. Due to the MTU restriction, the maximum payload size is $1500$ bytes, which means that each rule needs $1500$ encrypted patterns at most to match all position.
Note that the size of each encrypted pattern is a constant (see Fig.~\ref{fig:predicatevector}) without regarding the length of original pattern strings. 
Moreover, it is a tiny data structure: each encrypted pattern is only $23$ bytes (see Section~\ref{sec:eva}). 

%

The matching process is outlined in Algorithm~\ref{Alg2}. 
After uploading the encrypted pattern list to \MB, \GW~generates the encrypted traffic from the packet payload via $\MSHVE.\mathsf{Enc}$.
%
%
Later, \MB~uses the pattern list to check the traffic from \GW~via $\MSHVE.\mathsf{Query}$.
If a target packet includes a required pattern, \MB~can recover an action and apply it to the packet.

\begin{figure}[!t]
	\centering
	\includegraphics[width=\linewidth]{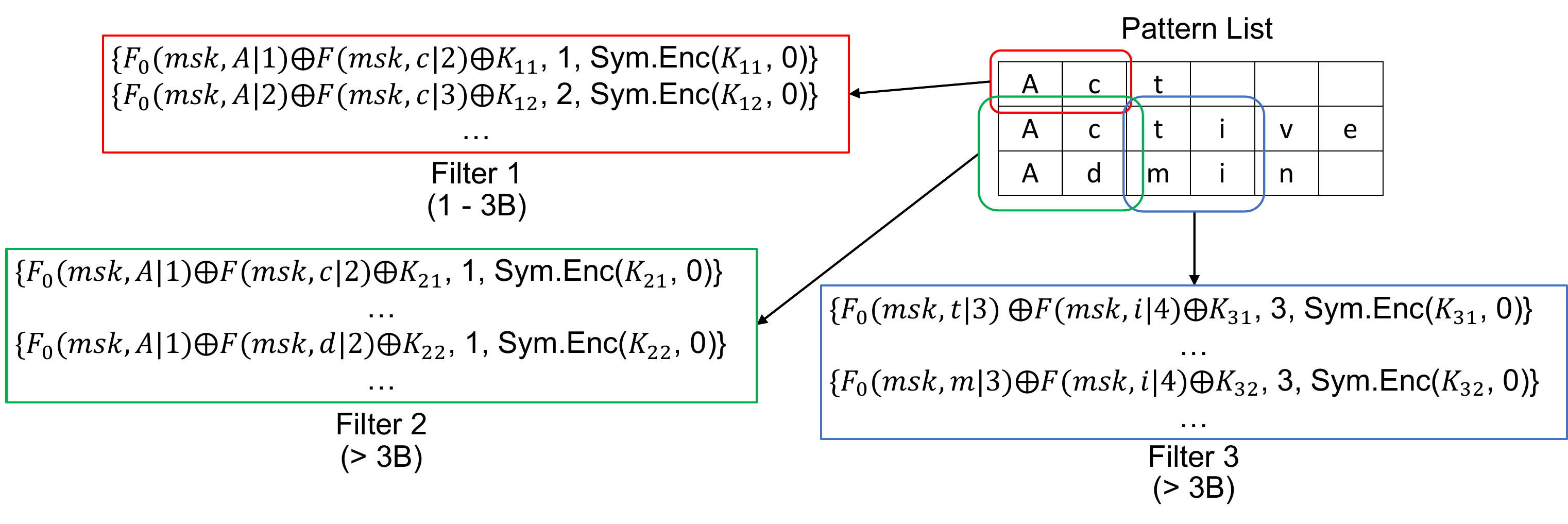}
	\caption{The proposed filter structure}
	\label{fig:filterstructure}
\end{figure}
{\bf\noindent Matching packet header and regular expression.} The protocol can be extended to support packet header inspection.
In particular, the header inspection focuses on the field information (e.g., HTTP header, HTTP method), which can only be found in the specific place in the header.
Thus, the protocol can parse the field information by extracting the field value as the pattern string and referring the header structure to compute the positional information.
Finally, the processed header inspection rule can be used by the original protocol to tackle with patterns that appear in the header.

For regular expression matching, it is common for the real-world pattern matching system like Snort to parse the regular expression as sub-strings and apply pattern matching algorithms to check those sub-strings respectively~\cite{choi2016dfc,stylianopoulos2017multiple}.
For instance, the regular expression ``ap*e'' aims to find the string start with `ap' and end with `e'.
The pattern matching system checks `ap' and `e' separately and returns match if the matching position of `e' is behind the one for `ap'.
Therefore, our protocol can follow the same strategy to check the regular expression for the encrypted traffic.
That is, the protocol generates encrypted patterns for `ap' and `e' separately and leverages the secret sharing scheme in~\cite{yuan2016privacy} to share the action into two encrypted patterns, and the action can only be recovered when two encrypted patterns are matched orderly.

{\bf\noindent \textbf{\textit{Remark}}.} The security properties of SHVE+ guarantee that the action can only be recovered if the encrypted payload includes the pattern (i.e. both the position and string should be matched).
%
Also, SHVE+ ensures that the equality of byte strings in the packet is not revealed to \MB, because the SHVE+ combines the packet payload and positions when generating the encrypted payload. 
Thus, the ciphertext of two identical bytes is different if they are in the different position of the payload.
A detailed security proof is given in Section~\ref{sec:security}.

Regarding the efficiency of the basic matching protocol, for each pattern, it performs a byte-to-byte match on the incoming traffic.
Recall that the protocol generates multiple encrypted patterns to match all specified starting positions of patterns in the traffic, its performance can be optimised via parallel processing.
In particular, the middlebox can use those independent encrypted patterns to perform pattern matching in specified positions concurrently on the encrypted traffic.
However, the drawback of this basic matching protocol is that its performance may degrade rapidly with the increasing size of the ruleset.
That is because each newly-added rule can have up to $1500$ more corresponding encrypted patterns.
It indicates that \MB~may need to perform $1500$ more $\MSHVE.\mathsf{Query}$ on a given packet if the size of the ruleset increased by one.
Next, we will introduce a secure filter to address the above issue.

\begin{algorithm}[!t]
	\caption{Encrypted Filter Generation}
	\label{Alg3}
	\small
	\begin{algorithmic}[1]
		\Require The master secret key $msk$ of SHVE; the ruleset $\mathit{R}$
		\Ensure  The encrypted filter $\mathcal{F}=\{\mathcal{F}_1, \mathcal{F}_2, \mathcal{F}_3\}$
		\Function {FilterGenerate}{$msk$, $\mathit{R}$}
		\State Parse $\mathit{R}$ as a pattern-action list $T=\{(pat, act)\}$
		\For{each $pat.\textrm{string}$ in $T$}
		\State $start\leftarrow pat.\textrm{offset}>0?pat.\textrm{offset}:1$
		\State $end\leftarrow pat.\textrm{depth}>0?start + pat.\textrm{depth}:1500$
		\State $s_1\leftarrow pat.\textrm{string}.substring(1, 2)$
		\For{$i=start : end - pat.\textrm{string}.len+1$}
		\State $\mathbf{v}_i\leftarrow *^{1500}$
		\State Insert $s_1$ at $\mathbf{v}_i[i]$
		\State $\mathbf{t}_i\leftarrow\SHVE.\mathsf{KeyGen}(msk, \mathbf{v}_i)$
		\If{$pat.\textrm{string}.len\leq 3$}
		\State Store $\mathbf{t}_i$ in $\mathcal{F}_1$
		\Else
		\State Store $\mathbf{t}_i$ in $\mathcal{F}_2$
		\State $s_2\leftarrow pat.\textrm{string}.substring(3, 2)$
		\State $j=i+2$
		\State $\mathbf{v}_j\leftarrow *^{1500}$
		\State Insert $s_2$ at $\mathbf{v}_j[j]$
		\State Store $\SHVE.\mathsf{KeyGen}(msk, \mathbf{v}_j)$ in $\mathcal{F}_3$
		\EndIf
		\EndFor
		\EndFor
		\State \Return $\mathcal{F}$
		\EndFunction
	\end{algorithmic}
\end{algorithm}
\subsection{Secure Filtering}
One key observation is that only a small fraction of the traffic includes malicious payloads (less than $0.01\%$ as shown in~\cite{choi2016dfc}).
Consequently, if we can efficiently distinguish the innocuous traffic from the malicious one, and only do pattern matching on the malicious traffic, the performance of the overall middlebox system can be highly improved.
To achieve our goal, we propose a secure filter system that can quickly evaluate whether the packet includes a match and where is the possible starting position to match.
%
Also, the filter is encrypted to prevent \MB~from learning any private information about the traffic and ruleset as in Section~\ref{sec:threat}.

The proposed filter consists of three filters in two-level (see Fig.~\ref{fig:filterstructure}).
%
%
The first level has two filters: Filter 1 stores information about the pattern strings that less than $4$ characters (bytes), while Filter 2 accounts for the longer patterns.
Both of them keep an encrypted pattern list of the beginning two bytes of each pattern string; it also combines the position information to check all position in the packet.
The filter in the second level (Filter 3) works together with Filter 2; it is a progressive filter generated from the next 2 bytes in the pattern string.
The progressive filter matches the following two bytes in each pattern string if it matched in Filter 2, and it reduces the false positive rate when matching a longer pattern.
Note that similar design philosophy is also adapted in plaintext traffic pattern matching systems~\cite{choi2016dfc, stylianopoulos2017multiple}.

Algorithm~\ref{Alg3} presents the steps of building the encrypted filter. 
For each pattern string, \GW~extracts the first two characters and generates encrypted patterns via $\SHVE$. Then, it inserts the encrypted patterns into either $\mathcal{F}_1$ or $\mathcal{F}_2$ by referring the length of the pattern string.
%
%
%
For those longer patterns (more than 3 bytes), the next two bytes are also generated as encrypted patterns and stored in $\mathcal{F}_3$.
Finally, \GW~uploads $\mathcal{F}$ with the above three sub-filters to \MB~as the encrypted filter.

To execute the secure filtering algorithm (cf. Algorithm~\ref{Alg4}), \MB~reuses the encrypted traffic to check the encrypted filter. 
In specific, as SHVE supports secure membership testing, \MB~is capable of recovering a ``True'' after $\SHVE.\mathsf{Query}$ if the upcoming payload has two bytes that match the ruleset pattern.
After applying the secure filtering, \MB~only requires to check the position that returns ``True'' when running the following pattern matching process.
Hence, the secure filtering can highly boost the overall pattern matching procedure, because for each rule, instead of using all encrypted patterns to check the whole encrypted traffic, only a few patterns corresponding to the filtered positions need to be checked.

\begin{algorithm}[!t]
	\caption{Secure Filtering}
	\label{Alg4}
	\small
	\begin{algorithmic}[1]
		\Require \GW~inputs the master secret key $msk$, the payload $P$; \MB~inputs the encrypted filter $\mathcal{F}$
		\Ensure  A list of possible matching positions $M$
		\Function {Filtering}{$msk$, $\mathcal{F}$, $\mathit{R}$}\\
		On \GW:
		\State Parse $P$ as a byte array and compute the encrypted traffic $\mathbf{c}\leftarrow\SHVE.\mathsf{Enc}(msk, P)$
		\State Send $\mathbf{c}$ to \MB
		\\
		On \MB:
		\For{i = 0 to $\mathcal{F}_1.len$}
		\If{$\SHVE.\mathsf{Query}(\mathcal{F}_{1}[i], \mathbf{c})=$``True''}
		\State Add $\mathcal{F}_{1}[i].S$ to $M$
		\EndIf
		\EndFor
		\If{$\mathbf{c}.len> 3$}
		\For{i = 0 to $\mathcal{F}_2.len$}
		\If{$\SHVE.\mathsf{Query}(\mathcal{F}_{2}[i], \mathbf{c})=$``True''}
		\For{j = 0 to $\mathcal{F}_3.len$}
		\If{$\SHVE.\mathsf{Query}(\mathcal{F}_{3}[j], \mathbf{c})=$``True''}
		\State Add $\mathcal{F}_{2}[i].S$ to $M$
		\EndIf
		\EndFor
		\EndIf
		\EndFor
		\EndIf
		\State \Return $M$
		\EndFunction
	\end{algorithmic}
\end{algorithm}
{\bf\noindent Filtering in parallel.} The secure filtering relies on two separate groups of filters (filters for pattern $\leq3$ bytes and $>3$ bytes).
Therefore, we can use the output to check the encrypted pattern corresponding to the pattern string $\leq3$ bytes and $>3$ bytes, respectively.
This can reduce the workload in the pattern matching process further because only the pattern that fits the size requirement needs to be checked after adopting this optimisation.
To achieve this, we slightly modify Algorithm~\ref{Alg4}: The matching positions output from $\mathcal{F}_1$ and $\mathcal{F}_3$ are kept in two matching position lists ($M_1$ and $M_2$).
Also, we employ two separate buckets to store the encrypted patterns for the pattern strings $\leq3$ bytes and $>3$ bytes separately.
As a result, \MB~can use the position information in $M_1$ to check the short patterns while utilising $M_2$ to check those longer patterns.

\section{Security Analysis}
\label{sec:security}
We give a security analysis to demonstrate that $\MB$ cannot learn the sensitive data in the ruleset as well as traffic during the pattern matching process.
We are the first to formalise the adversary capability in two aspects: 1) The adversary can select the packet to be challenged and get the encrypted patterns and filter selected by himself/herself. 
The goal of the adversary is to learn the sensitive data in the packet; 
2) The adversary can select the ruleset to be challenged and get the encrypted packet chosen by himself/herself.
The adversary aims to learn information about the ruleset other than the pattern matching result. 
Note that the existing work only considers either the security of the packet~\cite{sherry2015blindbox, fan2017spabox} or the security of the ruleset~\cite{yuan2016privacy, guo2018enabling}.

We follow the simulation-based security~\cite{lai2018result} to define a leakage function $\mathcal{L}$ for our encrypted pattern matching protocol $\mathcal{P}$ and then construct a simulator to show that $\mathcal{P}$ is $\mathcal{L}$-secure against adversaries as described in above.
More specifically, we construct a simulator $\mathcal{S}$ and prove that $\mathcal{S}$ can simulate $\mathcal{P}$ by using the leakage function $\mathcal{L}$ only.
This implies that the proposed protocol does not reveal any information about the packet payload and rules beyond the leakage function.

{\bf\noindent Security of SHVE+.} SHVE+ has a similar security model as SHVE~\cite{lai2018result} except that SHVE+ has a non-empty message space to support message encryption.
Recall that the security model of SHVE defines the {\it attribute-hiding} property, which indicates that the adversary can only learn two leakage functions: $\alpha(\mathbf{v})$ representing the wildcard pattern (positions) of a given predicate vector $\mathbf{v}$, and $\beta(\mathbf{v}, \mathbf{x})$ describing the leakage after queries on a given attribute vector $\mathbf{x}$ (i.e., leaking whether $\mathbf{v}$ and $\mathbf{x}$ are matched or not).
An adversary can arbitrarily request the SHVE trapdoors given the above two leakage functions to $\Sim_{\SHVE}$.
However, no more information about $\mathbf{v}$ and $\mathbf{x}$ will be leaked.
We keep the $\alpha(\mathbf{v})$ unchanged because the wildcard pattern of SHVE+ is exactly the same as in SHVE, while the definition of $\beta(\mathbf{v}, \mathbf{x})$ is modified as follows:
$\beta^{'}(\mathbf{v}, \mathbf{x})=m$ if $P_{\bf v}({\bf x})=1$, otherwise $\beta^{'}(\mathbf{v}, \mathbf{x})\xleftarrow{\$}\{0, 1\}^\lambda$.
The following theorem states the security of SHVE+:
\begin{theorem}\label{thm:MSHVE}
	SHVE+ is attribute-hiding in the ideal cipher model under the security model defined in above.
\end{theorem}
We omit the proof of Theorem~\ref{thm:MSHVE} because its statement and proof are exactly the same as Theorem 1 of~\cite{lai2018result}.
In the rest of this section, we directly apply the simulator of SHVE+ when simulating $\P$.

{\bf\noindent Security of the pattern matching protocol.} Let $\P$ be the pattern matching protocol. The security of $\P$ is formally defined via two real/ideal game definitions.
The first real/ideal game definition depicts the security of $\P$ against the adversary $\Adv_1$ who aims to compromise the confidentiality of the encrypted packets.
This adversary is identical to the one in~\cite{SherryHS12} who entrenches in \MB~and can get any number of the encrypted patterns and filter tokens to examine the encrypted packet.
We now describe the leakage function $\mathcal{L}_1$ towards $\Adv_1$.
On input a packet $P$ and the adaptively chose ruleset $R$, the leakage function can be parameterised as $\mathcal{L}_1(P,R)=\{\mathsf{Pos_f}, \mathsf{Pos_e}, \mathsf{Act}\}$ formed as follows:
\begin{itemize}
	\item $\mathsf{Pos_f}$ is the possible matching positions in $P$ w.r.t. $R$. Formally, $\mathsf{Pos_f}[i]$ is an array of possible matching positions in $P$ w.r.t. $R[i]$.
	\item $\mathsf{Pos_e}$ is the matched positions in $P$ w.r.t. $R$. Formally, $\mathsf{Pos_e}[i]$ is an array of matched positions in $P$ w.r.t. $R[i]$.
	\item $\mathsf{Act}$ is the actions that need to be performed on $P$. Formally, if $R[i]$ is matched in $P$, $\mathsf{Act}[i]=R[i].act$, otherwise, $\mathsf{Act}[i]=\bot$.
\end{itemize}
The following games and theorem state that $\P$ can protect the packet confidentiality in the presence of $\Adv_1$:
\begin{itemize}
	\item $\Real_{\Adv_1}^{\P} (\lambda)$: The adversary $\Adv_1$ chooses a packet $P$ to generate the encrypted packet $\mathbf{c}$ and gives $\mathbf{c}$ to $\Adv_1$. 
	Then, $\Adv_1$ adaptively chooses a rule $r$ to query. 
	To respond, the game runs $\textsc{Generate}(msk, r)$ and $\textsc{FilterGenerate}(msk, r)$.
	Later, the game gives the protocol outputs to $\Adv_1$. 
	Eventually, $\Adv_1$ outputs a bit.
	\item $\Ideal_{\Adv_1}^{\Sim} (\lambda)$: The game initialises a counter $i = 0$ and an empty list $R$.
	The adversary $\Adv_1$ chooses a packet $P$, and the game generates the encrypted packet $\mathbf{c}\leftarrow\Sim(\mathcal{L}_1(P))$ and gives $\mathbf{c}$ to $\Adv_1$. 
	Then, $\Adv_1$ adaptively chooses a rule $r$ to query. 
	To respond, the game records the rule as $R[i]$, gives the output of $\Sim(\mathcal{L}_1(P, R))$ ($R$ keeps all history rules) to $\Adv_1$ and increases $i$ by $1$. 
	Eventually, $\Adv_1$ outputs a bit.
\end{itemize}
\begin{theorem}\label{thm:S1}
	$\mathcal{P}$ is $\mathcal{L}_1$-secure against $\Adv_1$, assuming that the SHVE+ scheme is selectively simulation-secure, that the SHVE scheme is selectively simulation-secure.
\end{theorem}
\begin{IEEEproof}	
	We show that we could combine $\mathcal{L}_1$, $\Sim_{\SHVE}$ and $\Sim_{\MSHVE}$ to simulate $\P$.
	Suppose $\Adv_1$ provides a packet $P$ to $\Sim_1$. $\Sim_1$ invokes $\mathbf{c}\leftarrow\Sim_{\MSHVE}(\lambda)$ to generate the ciphertext of $P$ and gives it to $\Adv_1$.
	Upon receiving the $i$-th rule from $\Adv_1$, $\Sim_1$ refers to $\mathsf{Pos_f}[i]$ to simulate the filter. 
	In specific, for each possible matching position $p_f\in\mathsf{Pos_f}[i]$, $\Sim_1$ sets $\{p_f,p_f+1\}$ as $\alpha_{i1}(\mathbf{v}_{i1})$ and $\beta_{i1}(\mathbf{v}_{i1}, P)$ as ``True''.
	If $|R[i]|\geq 3$, $\Sim_1$ additionally sets $\{p_f+2,p_f+3\}$ as $\alpha_{i2}(\mathbf{v}_{i2})$ and $\beta_{i2}(\mathbf{v}_{i2}, P)$ as ``True''.
	Then, it runs $\Sim_{\SHVE}(\alpha_{i1}(\mathbf{v}_{i1}), \beta_{i1}(\mathbf{v}_{i1}, P))$ and $\Sim_{\SHVE}(\alpha_{i2}(\mathbf{v}_{i2}), \beta_{i2}(\mathbf{v}_{i2}, P))$ (if $|R[i]|\geq 3$) to get the corresponding filter.
	Similarly, for each matched position $p_e\in\mathsf{Pos_e}[i]$, $\Sim_1$ sets $\alpha_{i}(\mathbf{v}_{i})=\{p_e, ..., p_e + |R[i]| - 1\}$ and $\beta_{i}^{'}(\mathbf{v}_i, P)=\mathsf{Act}[i]$.
	$\Sim_1$ then calls $\Sim_{\MSHVE}(\alpha_{i}(\mathbf{v}_{i}), \beta_{i}^i(\mathbf{v}_{i}, P))$ to get the token for $R[i]$.
	$\Adv_1$ finally receives the simulated filter and encrypted pattern corresponding to $R[i]$.
	
	It is obvious that the simulated ciphertext is indistinguishable from the real ciphertext as it is computed via $\PRF$.
	Additionally, the security of SHVE and SHVE+ directly ensure that the simulated trapdoors for the filter and encrypted patterns are indistinguishable from the real trapdoors generated by $\Real_{\Adv_1}^{\P} (\lambda)$.
	Thus, it concludes that for every adversary $\Adv_1$, it has a negligible probability to learn more information from $P$ than the defined leakage function $\mathcal{L}_1$.
\end{IEEEproof}

Theorem~\ref{thm:S1} shows that the adversary cannot infer any information about the packet beyond $\mathcal{L}_1$ after receiving the encrypted pattern and filter.
It indicates that $\MB$~cannot know any information about a legitimate packet as it does not match any rule.
Meanwhile, to fulfil the requirement of pattern matching middleboxes, $\MB$~is allowed to learn the matching information and action on a malicious packet.
This enables the pattern matching middlebox to apply the inspection rule on the malicious packet efficiently. 

We also consider the adversary $\Adv_2$ who wants to learn unintended information from the ruleset, which captures the capacity of adversaries either in \MB~or endpoints.
Similar to the adversary in~\cite{yuan2016privacy}, $\Adv_2$ is able to use arbitrary packet payload to examine the ruleset deployed on \MB.
Let $\mathbf{q}[i], 1\leq i\leq |\mathbf{q}|$ be a query packet and $R[j], 1\leq j\leq |R|$ is a rule in the given ruleset. We have $\mathcal{L}_2(R, \mathbf{q})=\{\mathbf{r}, |\mathcal{E}|, |\mathcal{F}|, \mathsf{Match_P}, \mathsf{Match}, \mathsf{Act}, \mathsf{IP}\}$ towards $\Adv_2$ formed as follows:
\begin{itemize}
	\item $\mathbf{r}$ is an array storing the length of each rule, i.e., $\mathbf{r}[i]$ is the length of $R[i]$.
	\item $|\mathcal{E}|$ is the size of the encrypted pattern list.
	\item $|\mathcal{F}|$ is the size of the encrypted filter.
	\item $\mathsf{Match_P}$ is the possible match position pattern of each packet $\mathbf{q}[i]$ w.r.t. each rule $R[j]$, which is a bidimensional array: $\mathsf{Match_P}[i,j]$ is all positions in $\mathbf{q}[i]$ that probably match $R[j]$.

	\item $\mathsf{Match}$ is the matched position pattern of each packet $\mathbf{q}[i]$ w.r.t. each rule $R[j]$, which is a bidimensional array: $\mathsf{Match}[i,j]$ is all positions in $\mathbf{q}[i]$ that match $R[j]$.
	\item $\mathsf{Act}$ is the action pattern of each packet $\mathbf{q}[i]$ w.r.t. each rule $R[j]$, which is a bidimensional array.
	If $R[j]$ matches in $\mathbf{q}[i]$, $\mathsf{Act}[i,j]=R[j].act$, otherwise, $\mathsf{Act}[i,j]=\bot$.
	\item $\mathsf{IP}$ is the intersection pattern of any two rules $R[j], R[k], 1\leq j, k\leq |R|, j\neq k$ matched the packet $\mathbf{q}[i]$, which is a three dimensional array.
	Particularly, $\mathsf{IP}[i,j,k]$ stores the intersection position of $R[j], R[k]$.
\end{itemize}
The following games and theorem indicate the security guarantee on the ruleset in the presence of $\Adv_2$:
\begin{itemize}
	\setlength{\listparindent}{\parindent}
	\item $\Real_{\Adv_2}^{\P} (\lambda)$: The adversary $\Adv_2$ chooses a ruleset $R$ and a packet list $\mathbf{q}$.
	Then, the game runs $\textsc{Generate}(msk, \mathit{R})$ and $\textsc{FilterGenerate}(msk, \mathit{R})$ to generate the encrypted pattern $\mathcal{E}$ and filter $\mathcal{F}$. 
	Later, the game runs $\textsc{Filtering}(msk, \mathcal{F}, \mathbf{q}[i])$ and $\textsc{Match}(msk, \mathcal{E}, \mathbf{q}[i])$ for $1\leq i\leq |\mathbf{q}|$.	
	The ciphertext $\mathbf{c_i}$ of $\mathbf{q}[i]$ and pattern matching results are stored in $\mathbf{t}[i]$ as the transcript.
	The generated $\mathcal{E}$, $\mathcal{F}$ and $\mathbf{t}$ are given to $\Adv_2$.
	Eventually, $\Adv_2$ outputs a bit.
	\item $\Ideal_{\Adv_2}^{\Sim} (\lambda)$: The adversary $\Adv_2$ chooses a ruleset $R$ and a packet list $\mathbf{q}$.
	Then, the game runs $\{\mathcal{E},\mathcal{F},\mathbf{t}\}\leftarrow\Sim_2(\mathcal{L}_2(R, \mathbf{q}))$ and gives the outputs to $\Adv_2$. 
	Eventually, $\Adv$ outputs a bit.
\end{itemize}
\begin{theorem}\label{thm:S2}
	$\mathcal{P}$ is $\mathcal{L}_2$-secure against $\Adv_2$, assuming that the SHVE+ scheme is selectively simulation-secure, that the SHVE scheme is selectively simulation-secure.
\end{theorem}
\begin{IEEEproof}
	We show how to simulate $\P$ via $\mathcal{L}_2$, $\Sim_{\SHVE}$ and $\Sim_{\MSHVE}$.
	First, $\Sim_2$ initialises a bi-dimensional array $A$ to store the auxiliary information for the simulation.
	Then, $\Sim_2$ leverages the outputs from $\mathsf{IP}$ and $\mathsf{Match}$ to simulate the encrypted packet.
	In specific, for each query packet $\mathbf{q}[i], 1\leq i\leq |\mathbf{q}|$:
	\begin{enumerate}
		\item $\Sim_2$ generates an empty array $\mathbf{c_i}$.
		\item For each rule $R[j], 1\leq j\leq |R|$:
				
		$\bullet$ For each matched position $pos\in \mathsf{Match}[i, j]$ of $R[j]$, $\Sim_2$ checks whether $A[R[j], l], pos\leq l\leq pos + \mathbf{r}[j] - 1$ is set as the wildcard symbol.
		
		$\bullet$ If any of the above value in $A[R[j], l]$ is wildcard and $\forall\mathsf{IP}[i, j, k]=\emptyset, 1\leq k\leq |R|, k\neq j$, then $\Sim_2$ simulate a ciphertext $A[R[j], l]\xleftarrow{\$}\{0, 1\}^\lambda$.
		
		$\bullet$ Otherwise, if $\exists\mathsf{IP}[i, j, k]\neq\emptyset A[R[j], l]=A[R[k], l]$.
		
		$\bullet$ If $\mathbf{c_i}[l]$ is not set, $\Sim_2$ sets $\mathbf{c_i}[l]=A[R[j], l]$.
		
		\item For all empty entries in $\mathbf{c_i}$, $\Sim_2$ randomly generates a ciphertext and fills it into those empty entries.
	\end{enumerate}
	
	Next, $\Sim_2$ leverages the encrypted packet and the leakage functions to simulate the encrypted patterns, filter and transcript, for each encrypted packet $\mathbf{c_i}, 1\leq i\leq |\mathbf{q}|$:
	\begin{enumerate}
		\item $\Sim_2$ puts $\mathbf{c_i}$ into $\mathbf{t}[i]$ and sets $\mathbf{c_i}$ as $\mathbf{x}_i$.
		\item  For each rule $R[j], 1\leq j\leq |R|$:
		
		$\bullet$ $\Sim_2$ puts $\mathsf{Match}[i, j]$, $\mathsf{Match_P}[i, j]$ and $\mathsf{Act}[i, j]$ into $\mathbf{t}[i]$.

		$\bullet$ For each matched position $pos\in \mathsf{Match}[i, j]$ of $R[j]$, $\Sim_2$ sets $\alpha_i(\mathbf{v}_j)= \{pos, ..., pos + \mathbf{r}[j] - 1\}$ and $\beta_i^{'}(\mathbf{v}_j, \mathbf{x}_i)=\mathsf{Act}[i, j]$.
		
		$\bullet$ $\Sim_2$ calls $\Sim_{\MSHVE}(\alpha_i(\mathbf{v}_j), \beta_i^{'}(\mathbf{v}_j, \mathbf{x}_i))$ to get the corresponding encrypted pattern and put it into $\mathcal{E}$.
		
		$\bullet$ For each possible match position $pos_p\in \mathsf{Match_P}[i, j]$ of $R[j]$, $\Sim_2$ sets $\alpha_{i1}(\mathbf{v}_{j1})= \{pos_p, pos_p + 1\}$ and $\beta_{i1}(\mathbf{v}_{j1}, \mathbf{x}_i)=\mathrm{True}$; If $\mathbf{r}[j]\geq 3$, $\Sim_2$ also sets $\alpha_{i2}(\mathbf{v}_{j2})$ $= \{pos_p + 2, pos_p + 3\}$ and $\beta_{i2}(\mathbf{v}_{j2}, \mathbf{x}_i)=\mathrm{True}$.
		
		$\bullet$ Then, $\Sim_2$ calls $\Sim_{\SHVE}(\alpha_{i1}(\mathbf{v}_{j1}), \beta_{i1}(\mathbf{v}_{j1}, \mathbf{x}_i))$ and $\Sim_{\SHVE}(\alpha_{i2}(\mathbf{v}_{j2}), \beta_{i2}(\mathbf{v}_{j2}, \mathbf{x}_i))$ (if $\mathbf{r}[j]\geq 3$) to get the corresponding filter and put it into $\mathcal{F}$.
	\end{enumerate}
	Finally, $\Sim_2$ generates dummy HVE trapdoors to pad $\mathcal{E}$ and $\mathcal{F}$ to $|\mathcal{E}|$ and $|\mathcal{F}|$, respectively.
	
	Due to the security properties of SHVE and SHVE+, the adversary cannot distinguish the real and simulated $\mathcal{E}$ and $\mathcal{F}$.
	Moreover, the transcript $\mathbf{t}$ is also indistinguishable since the ciphertext is simulated under the ideal cipher model, and the query history under real and ideal games are identical.
	Even if the adversary uses the $\mathcal{E}$ and $\mathcal{F}$ to examine the ciphertext, the output result is also indistinguishable. 
	Therefore, $\Adv_2$ only has a negligible probability to learn more information than the defined leakage function $\mathcal{L}_2$ from the ruleset.
\end{IEEEproof}

Theorem~\ref{thm:S2} shows that the adversary cannot get information about the ruleset more than $\mathcal{L}_2$ after receiving the ciphertext of chosen packets.
This guarantees an untrusted \MB~cannot learn the ruleset with arbitrary legitimate packets.
On the other hand, as a part of the pattern matching middlebox requirement, the matching information and action can be revealed towards the malicious packet.
Hence, \MB~can still effectively inspect the packet and execute actions on malicious packets.

\begin{figure*}[!t]
	\centering
	\subfloat[Inspection latency]{\includegraphics[width=0.3\linewidth]{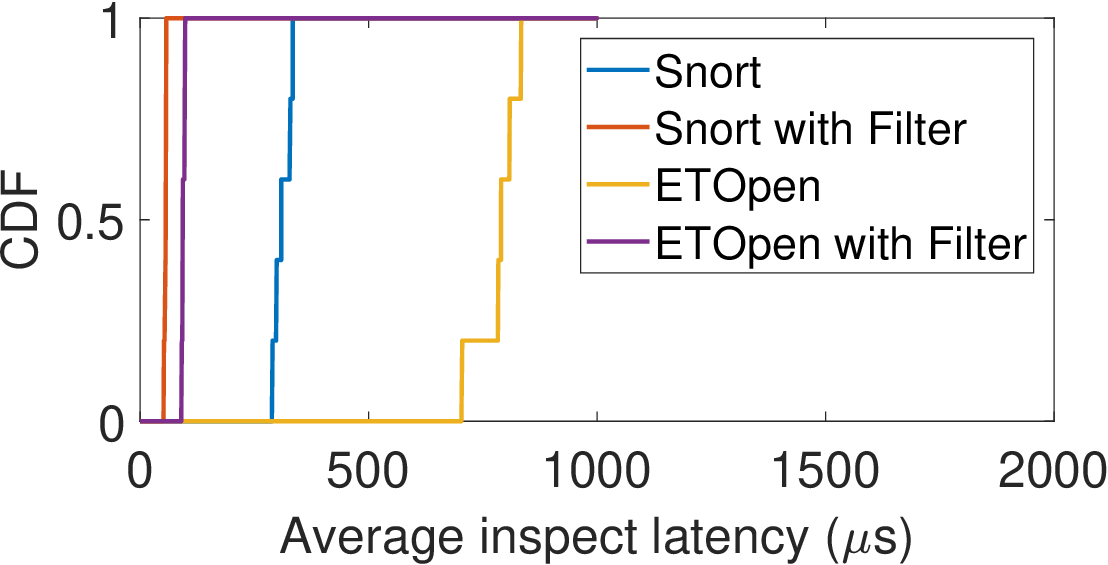}
		\label{fig:latency}}
	\hfil
	\subfloat[Bandwidth Overhead]{\includegraphics[width=0.3\linewidth]{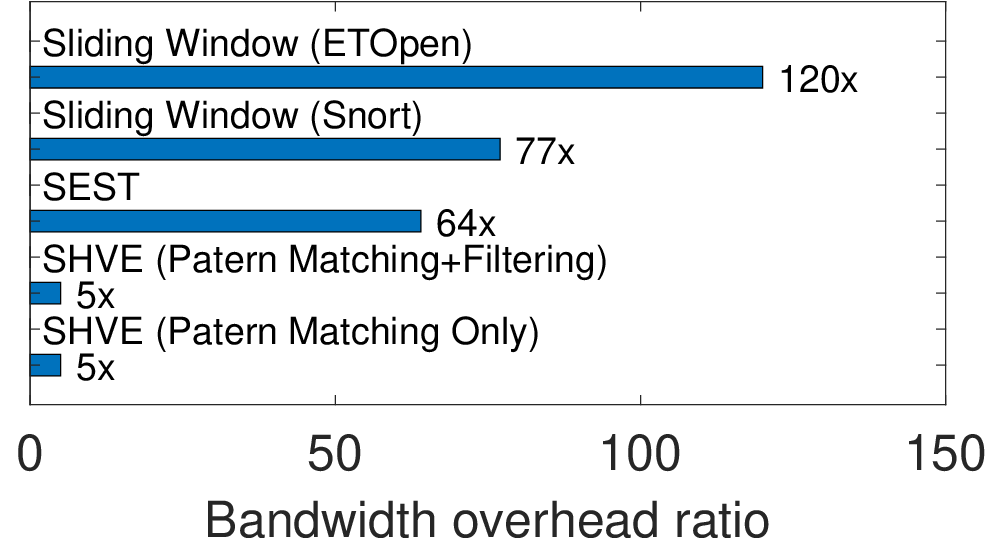}
		\label{fig:bandwidth}}
	\hfil
	\subfloat[Throughput]{\includegraphics[width=0.3\linewidth]{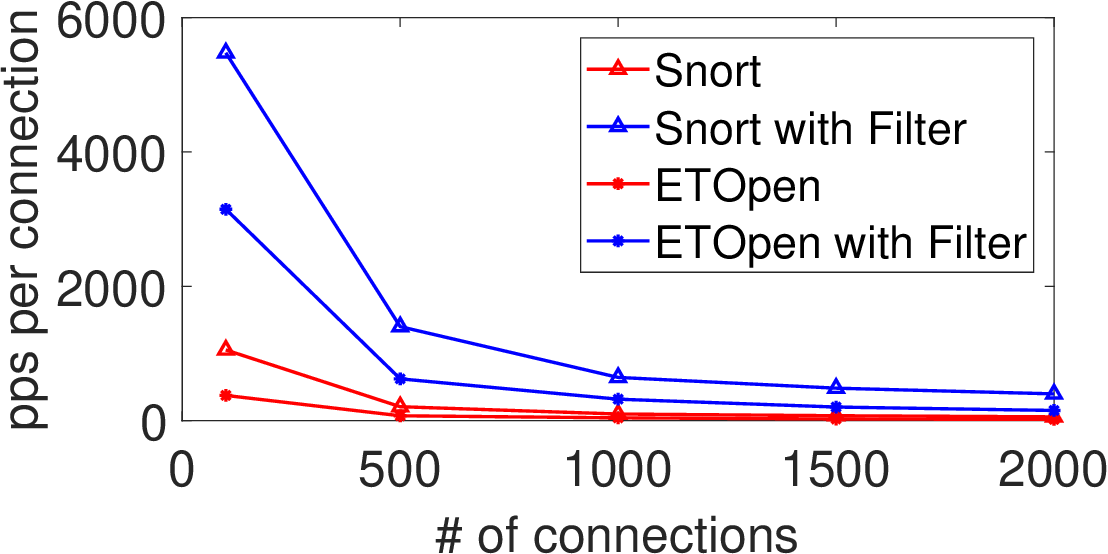}
		\label{fig:throughput}}
	\caption{The performance evaluation for the proposed middlebox module.} \label{fig:performance}
	\vspace{-15pt}
\end{figure*}

\section{Experiment and Evaluation}\label{sec:eva}
{\bf\noindent Environment setup and implementation.} We choose two open-source rulesets, i.e., Snort ruleset~\cite{snort2019ruleset} (1522 rules, 1116 patterns) and ETOpen ruleset\footnote{Emerging Threats ruleset: https://rules.emergingthreats.net} (24804 rules, 12634 patterns) to initialise our pattern matching middlebox module and use the traffic dump iCTF08\footnote{iCTF08 dumps: https://ictf.cs.ucsb.edu/archive/2008/dumps/} to evaluate its performance.

We implement the middlebox module and a gateway client in C++.
Recall that SHVE+ combines each pattern string with all possible positions to generate the encrypted pattern list. This treatment ensures that matching can only happen in the positions indicated in the rules. 
To save memory and bandwidth, we choose AES-CMAC as the PRF to implement SHVE and SHVE+ and truncate the output of PRF to 5 bytes as in~\cite{sherry2015blindbox,yuan2016bringing}.
Because the PRF outputs in the above scheme are used to mask the random key for the symmetric key encryption scheme $\Sym$, truncating them does not affect their correctness.
Hence, we use the PRF to mask a 5-byte random value and employ a KDF (Key Derivation Function) to generate the key for $\Sym$ from the random value.
Also, we substitute the non-wildcard position array $S$ (see Section~\ref{sec:MSHVE}) to a 2-byte integer value indicating the length of each pattern.
Note that this will not affect the correctness of SHVE+, because those positions represent a continuous string under the pattern matching application.
Finally, we stress that the truncation reduces the communication cost without affecting the security of our protocol.
That is because SHVE/SHVE+ trapdoors and ciphertexts are not decryptable as they consist of PRF values.
The adversary still cannot learn extra information about the packet and ruleset beyond the leakage profile in Section~\ref{sec:security}.

After optimisation, each encrypted pattern requires $23$ bytes  (1 PRF value + 1 AES ciphertext + 1 pattern length), and the encrypted pattern list for Snort ruleset costs $43.5$ MB, while the one for ETOpen requires $808$ MB.
%
On the other hand, the secure filtering protocol generates SHVE trapdoors for the beginning 2 bytes of each distinct pattern, and it further generates SHVE trapdoors for the following 2 bytes if the pattern is larger than 3 bytes.
We observe that the filtering protocol generates a $32$ MB filter from Snort ruleset, and $129$ MB for ETOpen ruleset.
These memory costs are moderate to a cloud server where the middlebox is supposed to deploy~\cite{jamshed2017mos,duan2019lightbox}.
In addition, computing and uploading the above lists are one-time costs in the initialisation phase, and it enables the middlebox to save bandwidth during the inspection phase tremendously.

For performance evaluation, we deploy the middlebox on a server equipped with Intel Core i7-6700 3.4GHz CPU and 16GB RAM and use a desktop with Intel Core i5-6500 3.2GHz CPU and 8GB RAM as the gateway client.

{\bf\noindent Performance evaluation.} First, we show the setup time for the middlebox module.
For the generation of the encrypted pattern list, it runs $\MSHVE.\mathsf{KeyGen}$ for each pattern and its all possible positions. This takes $18.9$ s in Snort ruleset and $287.3$ s in ETOpen ruleset.
Similarly, the filter generation combines each distinct 2 bytes extracted from the beginning of pattern with all possible positions and runs $\SHVE.\mathsf{KeyGen}$ operations to get the filter trapdoor, it also processes the next 2 bytes for the longer pattern. Our evaluation shows that it requires $14$ s and $45.5$ s for our two rulesets,  respectively.
%

Next, we report the runtime performance of the middlebox module.
In Fig.~\ref{fig:latency}, we evaluate the average inspection latency under two rulesets, respectively. 
For Snort ruleset, the inspection delay is less than $300~\mu$s. For the larger ETOpen ruleset, the inspection delay is less than $850~\mu$s.
We further examine the inspection latency after applying our secure filtering protocol. 
As a result, the middlebox only takes less than $60~\mu$s to inspect a packet in Snort ruleset, and $100~\mu$s for ETOpen ruleset, because in the iCTF08 dataset, only $1/6$ packets  and $1/9$ packets need further inspection against Snort ruleset and ETOpen ruleset, respectively.

As shown in Fig.~\ref{fig:bandwidth}, the bandwidth overhead in our proposed design is a constant, i.e., $5$ times in terms of the original packet size. 
This overhead is much smaller than any existing secure middlebox system using sliding window tokenisation algorithms~\cite{Desmoulins2018,sherry2015blindbox} because those algorithms enumerate all possible window sizes when tokenising the payload.
More specifically, for Snort ruleset, the number of distinct pattern sizes is $82$ (1 -- 214 bytes), and the sliding window tokenisation enlarges the bandwidth consumption by $77\times$.
%
For ETOpen ruleset, there are more distinct pattern sizes than Snort ruleset, i.e., $130$ (1 -- 196 bytes).
Therefore, the bandwidth overhead of the sliding window tokenisation reaches $120\times$.
In comparison, our system performs encryption and queries in byte-wise. Namely, it only scans the traffic once, and thus, the bandwidth overhead keeps constant, and it saves $94\%$ -- $96\%$ bandwidth comparing to the sliding window tokenisation in Snort and ETOpen ruleset.
Another approach (SEST~\cite{Desmoulins2018}) with constant bandwidth overhead is based on the elliptic curve.
However, the ciphertext in the elliptic curve is much longer than that in our symmetric building blocks, and this approach still leads to a prohibitive bandwidth overhead ($64\times$).
\begin{table}[!t]
	\centering
	\caption{Throughput of our middlebox for different rulesets.}
	\label{tbl:throughput}
	\tabcolsep 0.06in	
	\begin{tabular}{ccccc}
		\hline
		Ruleset & Snort & Snort (filter) & ETOpen & ETOpen (filter) \\
		\hline
		Throughput & $206$ MBps & $1442$ MBps & $75$ MBps & $578$ MBps \\
		\hline
	\end{tabular}
\end{table}

We simulate a multi-session scenario (100 to 2000 clients) to measure the throughput of the middlebox on our two different rulesets. 
The results are given in Fig.~\ref{fig:throughput} and Table~\ref{tbl:throughput}.
As our middlebox can perform filtering and matching efficiently in parallel, the throughput for each connection can reach up to $5000$ packets per second (pps) for $100$ connections under Snort ruleset, and around $300$ pps when there are $2000$ connections, and the overall throughput achieves $1442$ MBps.
For ETOpen ruleset, the throughput per connection still reaches $3000$ pps for $100$ connections, and the overall throughput is $578$ MBps.

\begin{table}
	\centering
	\begin{threeparttable}[!t]
		\caption{Theoretical performance comparison between the existing pattern matching middleboxes and our middlebox ($n$ is the packet size; $l$ is the size of pattern string).}
		\label{tbl:theoretical}
		\tabcolsep 0.06in	
		\begin{tabular}{cccc}
			\hline
			Scheme & Pattern size & Traffic size & Inspection time \\
			\hline 
			SEST~\cite{Desmoulins2018} & $1+\lceil l/8\rceil$ & $64n$ &$(n-l+1)T_{\mathsf{p}}\tnote{b}$\\
			\hline
			BlindBox~\cite{sherry2015blindbox} & $16\lceil l/16\rceil$ & \begin{tabular}[x]{@{}c@{}}$5((n+1)|D_{\mathsf{R}}\tnote{a}\ |$\\ $-\Sigma D_{\mathsf{R}})$ \end{tabular}& \begin{tabular}[x]{@{}c@{}}$((n+1)|D_{\mathsf{R}}|$\\ $-\Sigma D_{\mathsf{R}})T_{\mathsf{M}}$\tnote{c} \end{tabular}\\
			\hline
			Our middlebox & $23(n-l+1)$ & $5n$ &\begin{tabular}[x]{@{}c@{}} $(n-l+1)$\\ $(lT_{\mathsf{XOR}}\tnote{d}\ +T_{\mathsf{Dec}}\tnote{e}\ )$\end{tabular} \\
			\hline
		\end{tabular}
		\begin{tablenotes}
			\item [a] An array with all distinct pattern sizes in the ruleset
			\item [b] Time taken to compute a pairing
			\item [c] Time taken to access a tree index in memory
			\item [d] Time taken to compute an XOR operation
			\item [e] Time taken to decrypt a symmetric ciphertext
		\end{tablenotes}
	\end{threeparttable}
\end{table}
{\bf\noindent Comparison between prior designs.} We provide theoretical and real-world performance comparisons between SEST~\cite{Desmoulins2018}, BlindBox~\cite{sherry2015blindbox} and our middlebox. Note that the work\cite{yuan2016privacy, lan2016embark, fan2017spabox,ning2019privdpi} adopts a similar approach based on searchable encryption and tokenisation as BlindBox. Therefore, the comparison with BlindBox can also demonstrate our advantages to the above work.
The source code of~\cite{Desmoulins2018,sherry2015blindbox} is not publicly available, so we only compare our results with the one reported in their paper.
We note that the test machine of SEST has similar capabilities as ours, while BlindBox is evaluated on a much better machine (Intel Xeon E5-2650 2.6GHz, 128GB RAM).

In Table~\ref{tbl:theoretical}, we compare the theoretical performance from three perspectives of the listed works, i.e., the size of encrypted patterns, the size of the encrypted traffic ciphertext sending to the middlebox, and the inspection time on the middlebox.
The encrypted pattern size mainly affects the memory consumption of the middlebox. Although our scheme has the largest memory cost, it is still a moderate cost to a cloud server, as mentioned in the setup part.
On the other hand, the encrypted traffic size of our proposed scheme is much smaller than the other two schemes; it implies that our middlebox can save enormous bandwidth comparing with~\cite{Desmoulins2018,sherry2015blindbox}.
In terms of the inspection time, all schemes are linear in the length of the packet from the complexity view.
Nonetheless, the inspection time of our middlebox is comparable to BlindBox: both of them achieve a microsecond-level inspection delay because the inspection using the SHVE scheme is based on ultra-fast operations, i.e., XOR and $\Sym.\mathsf{Dec}$, which is only slightly slower than the index access operations in BlindBox.
However, the inspection delay of SEST is larger because it relies on cryptographic pairing, which can take a millisecond for each pairing.

\begin{table}[!t]
	\centering
	\caption{Performance comparison between the existing pattern matching middleboxes and the proposed middlebox using a $1500$-byte packet and Snort ruleset.}
	\tabcolsep 0.06in
	\label{tbl:real}	
	\begin{tabular}{ccc}
		\hline
		Scheme & Traffic size (bytes) & \begin{tabular}[x]{@{}c@{}} Inspection time\\ (1 rule (100 bytes), 1 packet) \end{tabular}\\
		\hline
		SEST~\cite{Desmoulins2018} & $96000$ & $600$ ms  \\
		\hline
		BlindBox~\cite{sherry2015blindbox} & $594700$ & $5~\mu$s \\
		\hline
		Our middlebox & $7500$ & $43~\mu$s \\
		\hline
	\end{tabular}
\end{table}
We report the performance comparison over real-world data in Table~\ref{tbl:real}.
We encrypt a 1500-byte ($n=1500$) packet as the encrypted traffic and use Snort ruleset ($|D_R|=82, \Sigma D_R=4142$) to inspect the traffic on the middlebox.
The result shows that our client only sends $7500$ bytes to the middlebox to inspect the given packet, which is $13$-$79$ times smaller than~\cite{Desmoulins2018,sherry2015blindbox}.
When inspecting a 100-byte pattern in the ruleset, although SEST~\cite{Desmoulins2018} and our middlebox leverage the linear scan to inspect the packet, SEST needs $600$ ms to finish the inspection  as it is based on the public-key cryptographic scheme, which is very slow in practice, while our middlebox based on SHVE only needs $43~\mu$s.
As reported in BlindBox~\cite{sherry2015blindbox}, inspecting one rule against a packet only requires $5~\mu$s. Note that our inspection delay is also in the microsecond-level, which is negligible in real-world scenarios.
Also, the testbed machine they used is much better than ours.

{\bf\noindent Deployment cost comparison.} To further illustrate the practicality of our middlebox, we estimate the deployment cost of our scheme and the representative (i.e., BlindBox~\cite{sherry2015blindbox}) of tokenisation-based approaches~\cite{sherry2015blindbox, yuan2016privacy, lan2016embark, fan2017spabox,ning2019privdpi,ning2020pine}.

Here, we assume that an enterprise deploys a 7/24 pattern matching middlebox on AWS to examine traffic, and a c5.2xlarge EC2 instance (8 cores, 16 GB RAM) is hired to host the middlebox module.
Note that this instance has sufficient memory for the proposed middlebox since our previous evaluation shows that $1$ GB is enough to store the encrypted pattern list and filter generated from ETOpen ruleset.
To have a consistent and stable network connection with higher bandwidth and throughput, the enterprise connects its network to the EC2 instance through AWS Direct Connect~\cite{aws2019direct}.
However, due to the traffic size under BlindBox is $15$ times larger than our middlebox, BlindBox needs higher network capacity in order to achieve similar performance as our middlebox.
For instance, if our middlebox requires $1$ Gbps bandwidth to guarantee a low delay, then BlindBox should use the $10$ Gbps plan instead.

\begin{table}[!t]
	\centering
	\caption{The hourly rate of memory and bandwidth under AWS pricing information.}
	\label{tlb:cost}	
	\begin{tabular}{|c|c|c|c|c|c|}
		\hline
		\multicolumn{4}{|c|}{Memory} & \multicolumn{2}{c|}{Bandwidth} \\
		\hline
		16 GB & 32 GB & 72 GB & 96 GB & 1 Gbps & 10 Gbps\\
		\hline
		$\$0.34$ & $\$0.68$ & $\$1.53$ & $\$2.04$ & $\$0.3$ & $\$2.25$ \\
		\hline
	\end{tabular}
\end{table}

\begin{table}[!t]
	\centering
	\caption{Monthly cost estimation between the tokenisation-based middleboxes and the proposed middlebox with ETOpen ruleset under AWS pricing information.}
	\label{tlb:deploy}	
	\tabcolsep 0.06in	
	\begin{tabular}{cccc}
		\hline
		Scheme & Instance & Network & Total\\
		\hline
		Blindbox~\cite{sherry2015blindbox} & $\$244.8$ & $\$1620$ ($10$ Gb bandwidth) & $\$1864.8$ \\
		\hline
		Our middlebox & $\$244.8$ & $\$216$ ($1$ Gb bandwidth) & $\$460.8$ \\
		\hline
	\end{tabular}
\end{table}

We refer to the pricing information in the U.S. East region (Virginia) to compute the price.
In Table~\ref{tlb:cost}, we provide the hourly rate of memory and bandwidth when deploying the middlebox in AWS.
The table illustrates that the average hourly cost on memory is around ($\$0.02$ per GB), which is $10\times$ cheaper than the bandwidth ($\$0.22$ to $\$0.3$ per Gbps).

Next, the monthly cost estimation result according to the above assumptions is listed in Table~\ref{tlb:deploy}.
The result shows that BlindBox has to pay $7.5\times$ more ($\$1620$ versus $\$216$) to get the same network performance as ours.
In total, the monthly cost of deploying our middlebox in AWS only needs $\$460.8$ while BlindBox takes $\$1864.8$, which indicates a $300\%$ extra cost.
From the result, it can also be seen that although our design introduces higher memory consumption than prior designs, it will not be an issue in deployment due to the low cost of memory in the cloud.

\section{Conclusion}
In this paper, we design a protocol that allows outsourced middleboxes to perform pattern matching over encrypted traffic.
We first design a customised SHVE scheme (SHVE+) and then build an encrypted pattern matching protocol based on it to protect network traffic and middlebox rules during the pattern matching process. 
Next, we design a secure filtering protocol to accelerate the pattern matching process.
We implement our protocol, and our evaluation on real-world rulesets and traffic dump illustrates its advantages in terms of bandwidth, inspection delay, throughput and deployment cost compared to the latest arts.

In future work, we will aim to reduce the memory cost of the SHVE scheme as our current solution incurs a heavy memory cost compared to existing cryptographic-based solutions~\cite{Desmoulins2018,sherry2015blindbox}.
Moreover, we will study how to extend our middlebox to support more complicated pattern matching process such as the regular expression with special characters and escaping.

\section*{Acknowledgment}
The authors would like to thank the anonymous reviewers for their valuable comments and constructive suggestions.
The work was supported in part by the Monash University Postgraduate Publications Award and the ARC Discovery Projects (DP180102199, DP200103308).

\balance

\bibliographystyle{IEEEtran}
\bibliography{reference}
\vspace{-30pt}
\begin{IEEEbiography}[{\includegraphics[width=1in,height=1.25in,clip,keepaspectratio]{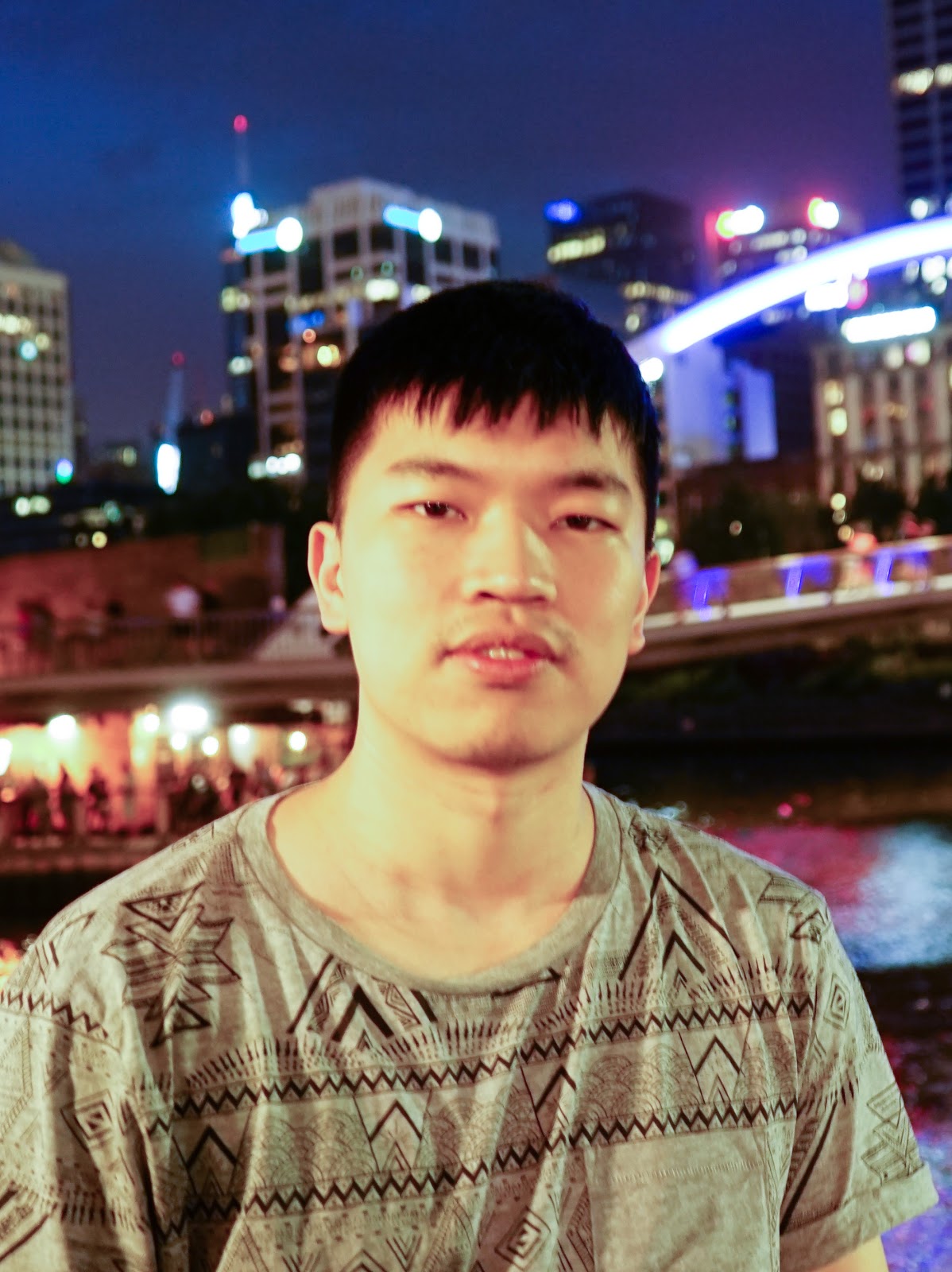}}]{Shangqi Lai} is a Research Fellow at the Faculty of Information Technology, Monash University. He got his PhD degree from Monash University in 2020. Besides, he received the B.E. degree from Nanjing University of Aeronautics and Astronautics in 2014, and the M.S. degree from The University of Hong Kong in 2015. His research interests include data privacy and cloud security.
\end{IEEEbiography}

\begin{IEEEbiography}[{\includegraphics[width=1in,height=1.25in,clip,keepaspectratio]{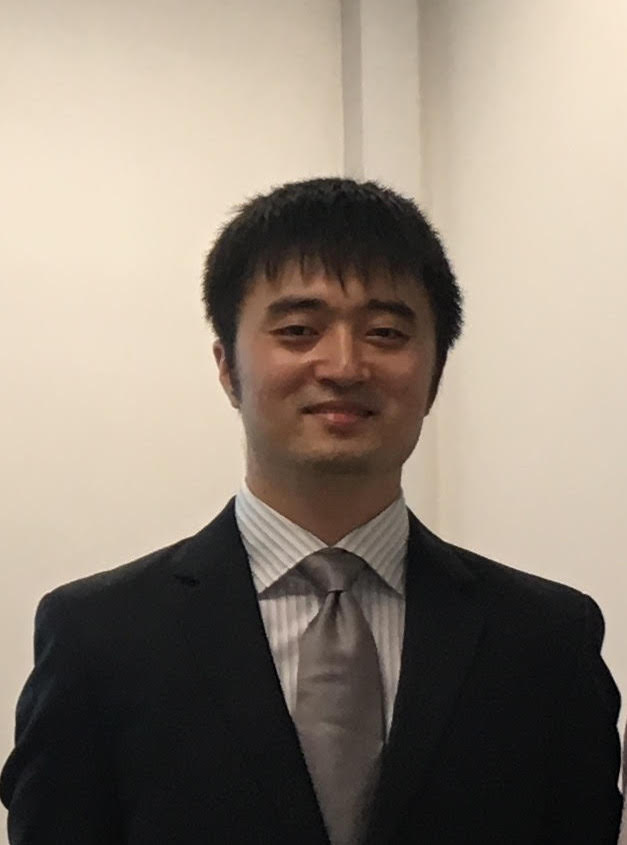}}]{Xingliang Yuan} received the B.S. and M.S. degrees in electrical engineering from the Illinois Institute of Technology, Nanjing University of Posts and Telecommunications and the Ph.D. degree in computer science from the City University of Hong Kong in 2016.

He is currently a Lecturer with the Faculty of Information Technology, Monash University, Australia. His research has been supported by Australian Research Council, CSIRO Data61, and Oceania Cyber Security Centre. His work has appeared in prestigious venues in security, computer networks, and distributed systems, such as ACM CCS, NDSS, ACM AsiaCCS, ESORICS, IEEE INFOCOM, IEEE ICDCS, IEEE ICNP, IEEE ICDE, IEEE Transactions on Dependable and Secure Computing, IEEE Transactions on Information Forensics and Security, IEEE/ACM Transactions on Networking, IEEE Transactions on Parallel and Distributed Systems, and IEEE Transactions on Mobile Computing. His research focuses on designing protocols and systems to address privacy and security issues in cloud and networked applications.
\end{IEEEbiography}

\vspace{-30pt}
\begin{IEEEbiography}[{\includegraphics[width=1in,height=1.25in,clip,keepaspectratio]{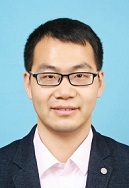}}]{Shi-Feng Sun} was awarded his Ph.D degree in Computer Science and Technology from Shanghai Jiao Tong University, China. During his Ph.D study, he worked as a visiting scholar in the Department of Computing and Information Systems at the University of Melbourne for one year. He was a postdoc fellow in Department of Computing at PolyU, Hong Kong under the supervision of A/Prof. Man Ho Au. After that, he worked as a Research Fellow in Cybersecurity group at Monash University. Currently, he is a Lecturer in the Faculty of Information Technology at Monash. His research interest centers on cryptography and data privacy, particularly on provably secure cryptosystems against physical attacks, data privacy-preserving technology in cloud storage, and privacy-enhancing technology in blockchain. He has published more than 40 quality papers, including publications in ACM CCS, NDSS, EUROCRYPT, PKC, ESORICS, AsiaCCS, and IEEE TDSC, etc.
\end{IEEEbiography}

\vspace{-40pt}
\begin{IEEEbiography}[{\includegraphics[width=1in,height=1.25in,clip,keepaspectratio]{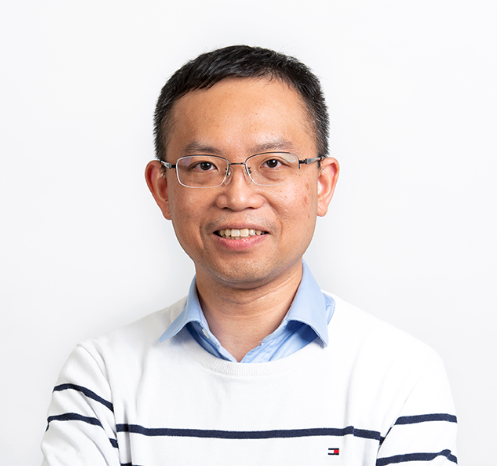}}]{Joseph K. Liu} is an Associate Professor in the Faculty of Information Technology, Monash University in Melbourne, Australia. He got his PhD from the Chinese University of Hong Kong at 2004. His research areas include cyber security, blockchain and applied cryptography. He has more than 200 publications in top venues such as CRYPTO, ACM CCS, NDSS, INFOCOM. He is currently the lead of the Monash Cyber Security Discipline Group. He has established the Monash Blockchain Technology Centre at 2019 and serves as the founding director. He has been given the prestigious ICT Researcher of the Year 2018 Award by the Australian Computer Society (ACS),  the largest professional body in Australia representing the ICT sector, for his contribution to the  cyber security community.
\end{IEEEbiography}

\nobalance
\vspace{-40pt}
\begin{IEEEbiography}[{\includegraphics[width=1in,height=1.25in,clip,keepaspectratio]{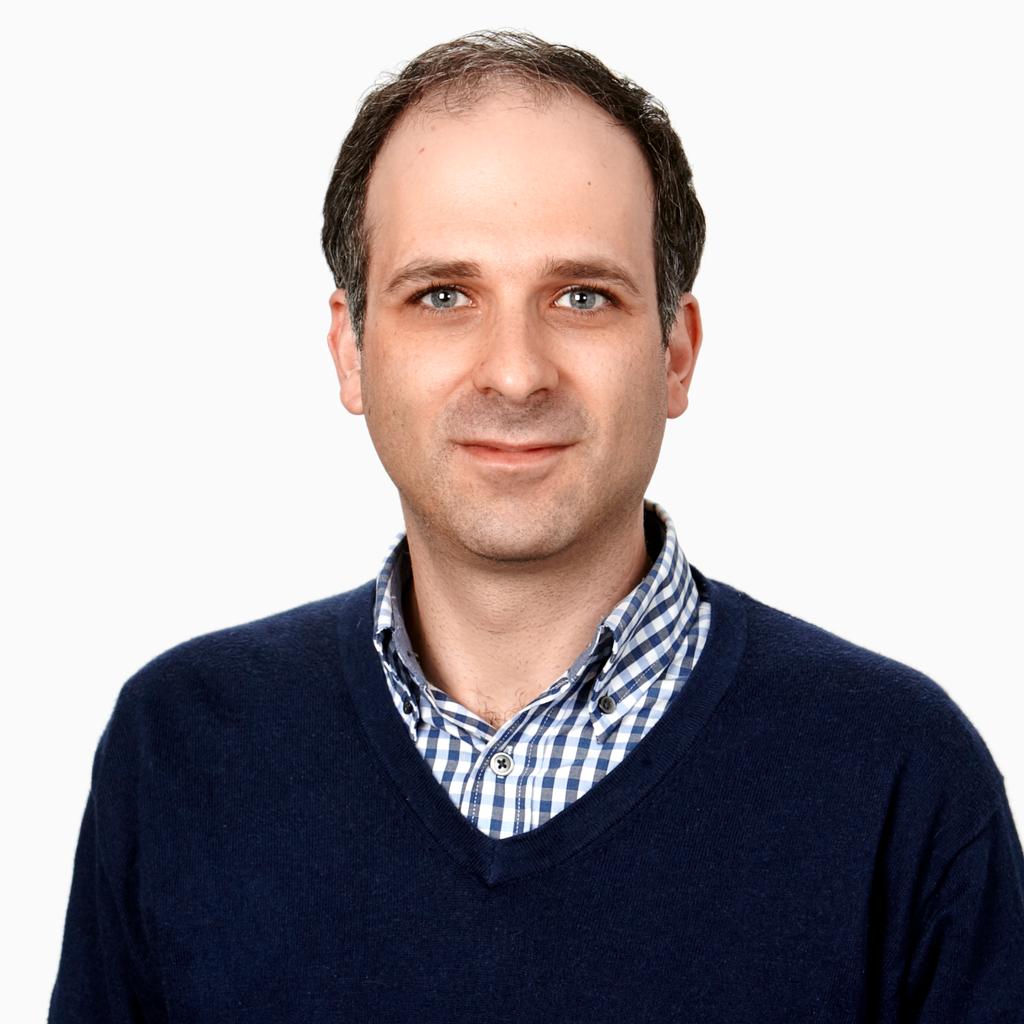}}]{Ron Steinfeld} is an Associate Professor at the Faculty of Information Technology, Monash University, Australia.  He received his Ph.D. degree in Computer Science in 2003 from Monash University, Australia, and has held a prestigious ARC Research Fellowship from 2009 to 2014. His main research interests are in the design and analysis of cryptographic algorithms and protocols, in particular in the areas of quantum-safe cryptography and secure computation protocols. He has over 20 years of research experience in cryptography and information security, and has published more than 70 research papers in international refereed conferences and journals, more than 10 of which have each been cited over 100 times. He received the ASIACRYPT 2015 best paper award for his contribution to lattice-based cryptography. He has served on the technical Program Committee of numerous international conferences in cryptography, is an editorial board member of the journal `Designs, Codes and Cryptography', and consults in cryptography design for the software industry.
\end{IEEEbiography}

\vspace{-40pt}
\begin{IEEEbiography}[{\includegraphics[width=1in,height=1.25in,clip,keepaspectratio]{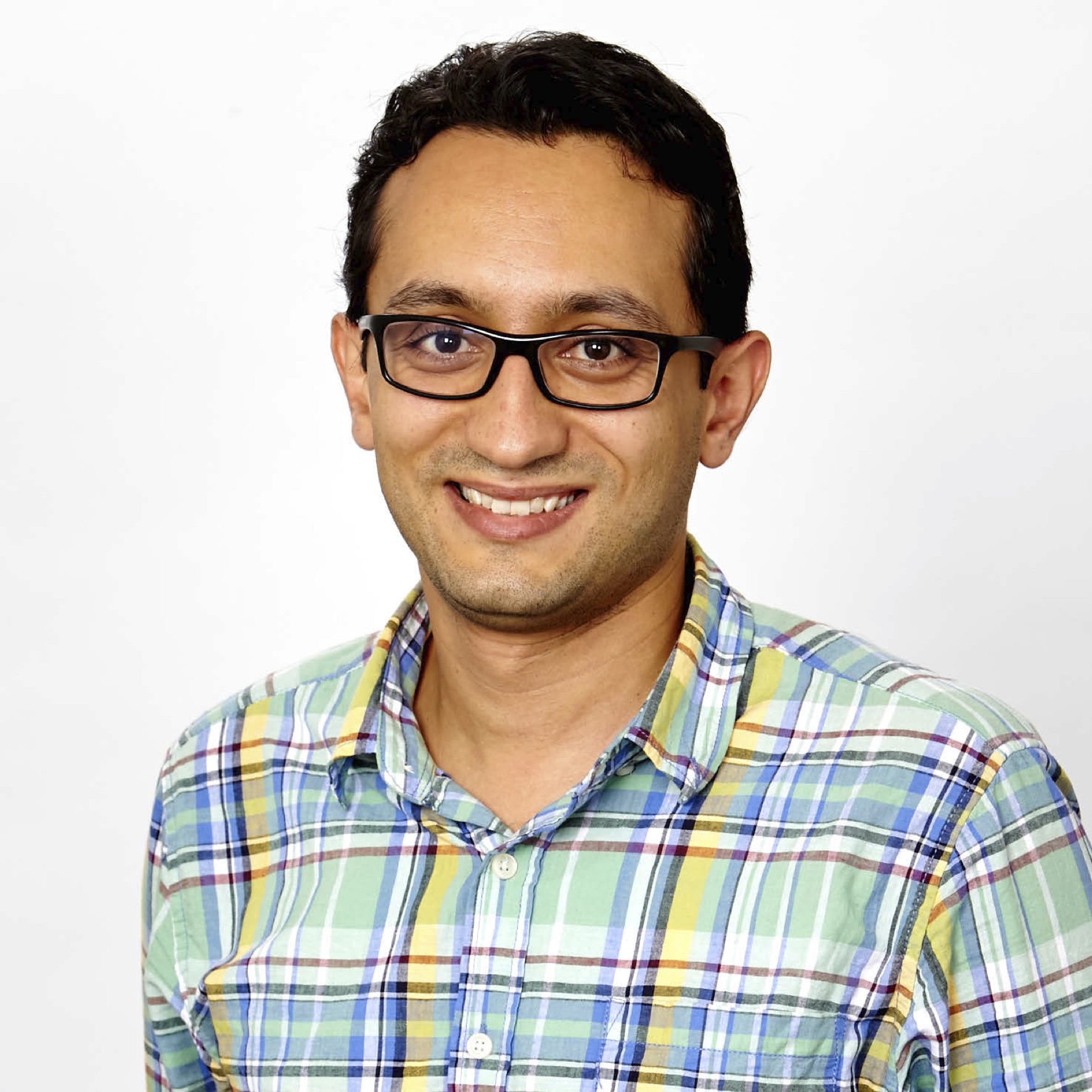}}]{Amin Sakzad} has got a Ph.D. degree in Applied Mathematics from Amirkabir University of Technology (AUT), Tehran, Iran, 2011. He was a research visitor and a lecturer at Carleton University, Ottawa, Canada, in 2010. He was a research lecturer at AUT in 2011. Starting from Jan 2012, he was a research fellow at Software-Defined Telecommunications (SDT) Laboratory in the Department of Electrical and Computer Systems Engineering at Monash University under the supervision of Prof. Emanuele Viterbo. From Feb 2015 to Apr 2017, he was a research fellow at Clayton School of Information Technology at Monash University under the supervision of Dr Ron Steinfeld. As of May 2017, he is a Lecturer (Assistant Professor) at the Faculty of Information Technology at Monash University. Dr. Amin Sakzad is mainly interested in applications of Euclidean lattices in cryptography and wireless communications. This includes applications of Algebraic Number Theory, Diophantine Approximation and Finite Fields in physical layer network coding and security, multiple-input multiple-output (MIMO) channels, lattice-based cryptography, and searchable encryption.
\end{IEEEbiography}

\vspace{-430pt}
\begin{IEEEbiography}[{\includegraphics[width=1in,height=1.5in,clip,keepaspectratio]{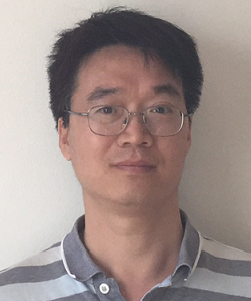}}]{Dongxi Liu} is a Senior Research Scientist in CSIRO since 2008. His research interests include applied cryptography, light weight encryption, and system security. His recent work aims to design public key encryption based on checkable hardness facts and design new proof-of-work blockchain protocol for crowdmining.
\end{IEEEbiography}

\end{document}